\documentclass[12pt,tightenlines,a4paper,twoside,altaffilletter,superscriptaddress,
               secnumarabic,showpacs,floatfix,nofootinbib,prd]{revtex4}

\usepackage[english]{babel}
\usepackage{amsmath,amssymb}
\usepackage{slashed}
\usepackage{epsfig}
\usepackage{graphicx,color}
\usepackage[dvips,paper=a4paper,inner=25mm,outer=19mm,
            top=42mm,bottom=20mm]{geometry}

\makeatletter
\def\NAT@sort{0}
\renewcommand{\NAT@sort@cites}[1]{\edef\NAT@cite@list{#1}}
\makeatother

\newcommand{\bra}[1]{\ensuremath{\langle #1 |}}   
\newcommand{\ket}[1]{\ensuremath{| #1 \rangle}}   

\renewcommand{\vec}[1]{{\mathbf{#1}}}
\newcommand{\MB}{M\"{o}ssbauer}
\newcommand{\Hamil}{H}
\renewcommand{\H}{{\rm H}}
\newcommand{\He}{{\rm He}}
\newcommand{\tp}{{\ensuremath{t_1}}}
\newcommand{\tpp}{{\ensuremath{t_2}}}
\newcommand{\tppp}{{\ensuremath{t_3}}}
\definecolor{grey}{rgb}{0.5,0.5,0.5}

\begin{document}

\title{Oscillations of \MB\ neutrinos}
\author{Evgeny Kh.~Akhmedov} \email[Email: ]{akhmedov@mpi-hd.mpg.de}
\affiliation{Max--Planck--Institut f\"ur Kernphysik, \\
             Postfach 10 39 80, 69029 Heidelberg, Germany \\[0.3cm]}
\affiliation{National Research Centre Kurchatov Institute, \\
             Kurchatov Sq.~1, 123182 Moscow, Russia \\[-0.3cm] \ }
\author{Joachim Kopp}        \email[Email: ]{jkopp@mpi-hd.mpg.de}
\author{Manfred Lindner}     \email[Email: ]{lindner@mpi-hd.mpg.de}
\affiliation{Max--Planck--Institut f\"ur Kernphysik, \\
             Postfach 10 39 80, 69029 Heidelberg, Germany \\[0.3cm]}
\date{\today}

\begin{abstract}
We calculate the probability of recoilless emission and detection of neutrinos 
(\MB\ effect with neutrinos) taking into account the boundedness of the parent 
and daughter nuclei in the neutrino source and detector as well as the 
leptonic mixing. We show that, in spite of their near monochromaticity, the 
recoillessly emitted and captured neutrinos oscillate. After a qualitative
discussion of this issue, we corroborate and extend our results by computing
the combined rate of $\bar{\nu}_e$ production, propagation and detection in
the framework of quantum field theory, starting from first principles. 
This allows us to avoid making any a priori assumptions about the energy and 
momentum of the intermediate-state neutrino. Our calculation permits 
quantitative predictions of the transition rate in future experiments, and 
shows that the decoherence and delocalization factors, which could in
principle suppress neutrino oscillations, are irrelevant under realistic
experimental conditions.
\end{abstract}

\maketitle

\section{Introduction}

Soon after the discovery of recoil-free emission and absorption of gamma rays
by \MB\ in 1958~\cite{Moessbauer:1958,Frauenfelder:1962},
it has been suggested by Visscher that a similar effect should also exist for
neutrinos emitted in electron capture processes from unstable nuclei embedded
into a crystal lattice~\cite{Visscher:1959}. In the 1980's, the idea was
further developed by Kells and Schiffer~\cite{Kells:1983,Kells:1984nm},
who showed that bound state beta decay~\cite{Bahcall:1961} could
provide an alternative recoilless production mechanism. In this case, an
antineutrino with a very small energy uncertainty would be emitted, which
could then be absorbed through induced orbital electron
capture~\cite{Mikaelyan:1967}. Recently, there has been a renewed interest
in this idea, inspired by two works by Raghavan~\cite{Raghavan:2005gn,
Raghavan:2006xf}, in which the feasibility of an experiment using the 
emission process
\begin{equation}
  {}^3\H \ \rightarrow \ {}^3\He~+~e^- \text{(bound)}~+~\bar{\nu}_e  
  \label{eq:prod}
\end{equation}
and the detection process 
\begin{equation}
  {}^3\He~+~e^- \text{(bound)}~+~\bar{\nu}_e \ \ \rightarrow  \ ^3\H
  \label{eq:abs}
\end{equation}
has been studied. The $^3\H$ and $^3\He$ atoms were proposed to be embedded
into metal crystals. The detection process would then have a resonance nature, 
leading to an enhancement of the detection cross section by up to a factor of 
$10^{12}$ compared to the non-resonance capture of neutrinos of the same 
energy. If such an experiment were realized, it could carry out a
very interesting physics program, including neutrino detection with 100 g  
scale (rather than ton or kiloton scale) detectors, searching for neutrino 
oscillations driven by the mixing angle $\theta_{13}$ at a baseline of 
only 10~m, determining the neutrino mass hierarchy without using matter 
effects, searching for active-sterile neutrino oscillations and studying the 
gravitational redshift of neutrinos~\cite{Raghavan:2005gn,Raghavan:2006xf,
Minakata:2006ne,Minakata:2007tn}. 

In this paper we consider recoillessly emitted and captured neutrinos -- which 
we will call \MB\ neutrinos -- from a theoretical point of view. In our 
discussion, we will mainly focus on the $^3\H$\,--$^3\He$ system of 
Eqs.~\eqref{eq:prod} and \eqref{eq:abs}, but most of our results apply also 
to other emitters and absorbers of \MB\ neutrinos.

One of our main goals is to resolve the recent controversy about the question
of whether \MB\ neutrinos would oscillate. It has been 
argued~\cite{Bilenky:2006hk} 
that the answer to this question depends on whether equal energies or equal 
momenta are assumed for different neutrino mass eigenstates -- the assumptions 
often made in deriving the standard formula for the oscillation probability.  
Moreover, a possible inhibition of oscillations due to the time-energy 
uncertainty relation has been brought up~\cite{Bilenky:2007vs}. 
To come to definitive conclusions regarding the oscillation phenomenology 
of \MB\ neutrinos, we employ a quantum field theoretical (QFT) approach, 
in which neutrinos are treated as intermediate states in the combined 
\mbox{production -- propagation -- detection} process and no {\em a priori} 
assumptions on the energies or momenta of the different neutrino mass 
eigenstates are made.

We begin in Sec.~\ref{sec:qualitative} by qualitatively discussing how the 
peculiar features of \MB\ neutrinos, and in particular their very small
energy uncertainty, affect the oscillation phenomenology. We argue that
oscillations do occur, and that the coherence length is infinite if line
broadening is neglected. We then proceed to quantitative arguments in
Sec.~\ref{sec:QM} and discuss a formula for the $\bar{\nu}_e$
survival probability in the quantum mechanical intermediate wave packet
formalism~\cite{Giunti:1997wq}, in which the neutrino is described as a
superposition of three wave packets, one for each mass eigenstate.
In Sec.~\ref{sec:QFT}, we derive our main result, the rate for the 
combined process of neutrino production, propagation and detection
in the QFT external wave packet approach. In this framework, 
the neutrino is described by an internal line in a Feynman 
diagram, while its production and detection partners are described by 
wave packets. Also in this section, for the first time, we calculate the 
rates for beta decay with production of a bound-state electron and for 
the inverse process of stimulated electron capture in the case of nuclei
bound to a crystal lattice. We distinguish between different neutrino line
broadening mechanisms and concentrate on the oscillation phenomenology, 
paying special attention to the coherence and localization terms in the 
$\bar{\nu}_e$ survival probability and to the \MB\ resonance conditions 
arising in each case. In Sec.~\ref{sec:discussion}, we discuss the obtained 
results and draw our conclusions.

\section{\MB\ neutrinos do oscillate}
\label{sec:qualitative}

\MB\ neutrinos have very special properties compared to those of neutrinos 
emitted and detected in conventional processes. In particular, they are almost 
monochromatic because they are produced in two-body decays of nuclei embedded 
in a crystal lattice and no phonon excitations of the host crystal accompany 
their production, which ensures the recoilless nature of this process. 
Therefore the width of the 
neutrino line is only limited by the natural linewidth, which is the 
reciprocal of the mean lifetime of the emitter, and by solid-state effects, 
including electromagnetic interactions of the randomly oriented nuclear spins,
lattice defects and impurities~\cite{Raghavan:2006xf,Potzel:2006ad,
Coussement:1992,Balko:1997}. 
For $^3\H$ decay, the natural linewidth is $1.17 \cdot 10^{-24}$~eV, but it
has been estimated that various broadening effects degrade this value 
to an experimentally achievable \MB\ linewidth of $\gamma =  
\mathcal{O}(10^{-11} \ \text{eV})$~\cite{Potzel:2006ad,Coussement:1992}.
Compared to the neutrino energy in bound state $^3\H$ decay, $E = 18.6$~keV,
the achievable relative linewidth is therefore of order $10^{-15}$.

In the standard derivations of the neutrino oscillation formula it is 
often assumed that the different neutrino mass eigenstates composing the 
produced flavor eigenstate have the same momentum ($\Delta p=0$), while  
their kinetic energies differ by $\Delta E\simeq \Delta m^2/2 E$. For 
bound state tritium beta decay (\ref{eq:prod}) and $\Delta m^2 =\Delta 
m_{31}^2\simeq 2.5\times 10^{-3}$ eV$^2$ one has $\Delta E\simeq 
7\times 10^{-8}$ eV, which is much larger than $\gamma$. 
One may therefore wonder if the extremely small energy uncertainty of 
\MB\ neutrinos would inhibit oscillations by destroying the coherence of 
the different mass eigenstates of which the produced $\bar{\nu}_e$
is composed. Indeed, if neutrinos are emitted with no momentum 
uncertainty and their energy uncertainty ($\sim\gamma$) is much smaller 
than the energy differences of the different mass eigenstates, in each  
decay event one would exactly know which mass eigenstate has been emitted.
This would prevent a coherent emission of different mass eigenstates, thus
destroying neutrino oscillations. If, on the contrary, one adopts the 
same energy assumption, the momenta of different mass eigenstates would 
differ by $\Delta p\simeq \Delta m^2/2p$, which would not destroy 
their coherence provided that the momentum uncertainty of the 
emitted neutrino state is greater that $\Delta p$; in that case, 
oscillations are possible.

It is well known that in reality neither same momentum nor same energy 
assumptions are correct 
\cite{Winter:1981kj,Giunti:1991ca,Giunti:2000kw,Giunti:2001kj,Giunti:2003ax}; 
however, for neutrinos from conventional sources both lead to the correct 
result, the reason being that neutrinos are ultra-relativistic and the spatial 
size of the corresponding wave packets is small compared to the oscillation 
length.%
\footnote{It is also essential that the energy and momentum 
uncertainties of these neutrinos are of the same order. }
The above assumptions are thus just shortcuts which allow one to 
arrive at 
the correct result in an easy (though not rigorous) way.  However, \MB\ 
neutrinos represent a very peculiar case, which requires a special 
consideration. 

Let us discuss the issue of coherence of different mass eigenstates in more 
detail. If one knows the values of the neutrino energy $E$ and momentum $p$ 
with uncertainties $\sigma_E$ and $\sigma_p$, from the energy-momentum relation 
of relativistic particles $E^2=p^2+m^2$ one can infer the value of the squared 
neutrino mass $m^2$ with the uncertainty 
\begin{align}
  \sigma_{m^2} = \sqrt{(2 E \sigma_E)^2 + (2 p \sigma_p)^2}\,,
  \label{eq:coherence-cond}
\end{align}
where it is assumed that $\sigma_E$ and $\sigma_p$ are independent. By 
$\sigma_E$ and $\sigma_p$ we will now understand the intrinsic quantum 
mechanical uncertainties of the neutrino energy and momentum, beyond which 
these quantities cannot be measured in a given production or detection process; 
$\sigma_{m^2}$ is then the quantum mechanical uncertainty of the inferred 
neutrino squared mass. A generic requirement for coherent emission of 
different mass eigenstates is their indistinguishability: the uncertainty 
$\sigma_{m^2}$ has to be larger than the mass squared difference $\Delta m^2$ 
~\cite{Kayser:1981ye}. From the above discussion, we know that for \MB\ 
neutrinos corresponding to the $^3\H$\,--$^3\He$ system one has $E 
\sigma_E \sim 10^{-8}$~eV$^2$, which is much smaller than $\Delta m^2\sim 
10^{-3}$ eV$^2$. Thus, whether or not \MB\ neutrinos oscillate depends on 
whether or not $2p\sigma_p>\Delta m^2$. 

While the energy of \MB\ neutrinos is very precisely given by the production 
process itself, this is not the case for their momentum. The neutrino momentum 
can in principle be determined by measuring the recoil momentum of the crystal 
in which the emitter is embedded. The ultimate uncertainty $\sigma_p$ of this 
measurement is related to the coordinate uncertainty $\sigma_x$ of the 
emitting nucleus through the Heisenberg relation $\sigma_p \sigma_x\ge 1/2$. 
Therefore, for the momentum uncertainty to be small enough to destroy the 
coherence of different mass eigenstates, $2p\sigma_p < \Delta m^2$, the 
coordinate uncertainty of the emitter must satisfy $\sigma_x \gtrsim 2p/\Delta 
m^2$. This means that the emitter should be strongly  de-localized with the 
coordinate uncertainty $\sigma_x$ of order of the neutrino oscillation 
length $L^{\rm osc}=4\pi p/\Delta m^2\simeq 20$~m. This is certainly not 
the case, because the coordinate uncertainty of the emitter cannot exceed 
the size of the source, i.e.~a few cm. In fact, it is even much smaller, 
because in principle it is possible to find out which particular nucleus 
has undergone the \MB\ transition by destroying the crystal and checking which 
$^3$H atom has been transformed into $^3$He. Thus, $\sigma_x$ is of the order 
of interatomic distances, i.e.~$\sigma_p\sim 10$ keV, so that 
\begin{equation}
2p\sigma_p \gg \Delta m^2\,.
\label{eq:local}
\end{equation}
This means that \MB\ neutrinos will oscillate. 
The condition (\ref{eq:local}) is often called the localization 
condition, because it requires the neutrino source to be localized in a 
spatial region that is small compared to the neutrino oscillation length 
$L^{\rm osc}$. 

It should be noted that for the observability of neutrino oscillations the 
coherence of the emitted neutrino state is not by itself sufficient; in 
addition, this state must not lose its coherence until the neutrino is 
detected. A coherence loss could occur because of the wave packet separation. 
When a neutrino is produced as a flavour eigenstate, the wave packets of its 
mass eigenstate components fully overlap; however, since they propagate with  
different group velocities, after a time $t^{\rm coh}$ or upon propagating a 
distance $L^{\rm coh}\simeq t^{\rm coh}$, these wave packets separate to 
such an extent that they can no longer interfere in the detector, and 
oscillations become unobservable. The coherence length $L^{\rm coh}$ depends 
on the energy uncertainty $\sigma_E$ of the emitted neutrino state and 
becomes infinite in the limit $\sigma_E\to 0$. 

{}From the above discussion it follows that the oscillation phenomenology 
of \MB\ neutrinos should mainly depend on their momentum uncertainty, 
whereas their energy uncertainty, though crucial for the \MB\ resonance 
condition, plays a relatively minor role for neutrino oscillations. 
Therefore, the equal energy assumption, though in general incorrect, 
should be a good approximation when discussing oscillations of \MB\ neutrinos. 
Adopting this approach, i.e.~assuming the neutrino energy to be \emph{exactly} 
fixed at a value $E$ by the production process, one obtains
for the $\bar{\nu}_e$ survival probability $P_{ee}$ at a distance $L$
\begin{align}
  P_{ee}
    &= \sum_{j,k} |U_{ej}|^2 |U_{ek}|^2
       \exp\Big[ -2\pi i \frac{L}{L^{\rm osc}_{jk}} \Big].
  \label{eq:infinite-wp-P1}
\end{align}
Here $U$ is the leptonic mixing matrix, $L^{\rm osc}_{jk}$ are the 
partial oscillation lengths, 
\begin{align}
  L^{\rm osc}_{jk} &= \frac{4\pi E}{\Delta m_{jk}^2}\,,
  \label{eq:Losc}
\end{align}
and the neutrinos are assumed to be ultra-relativistic or nearly 
mass-degenerate, so that 
\begin{align}
  \frac{\Delta m_{jk}^2}{2 E} \ll E \,.
  \label{eq:relativistic-approx}
\end{align}
Eq.~(\ref{eq:infinite-wp-P1}) is just the standard result for the $\bar{\nu}_e$ 
survival probability. As expected, we do not obtain any decoherence factors 
if the neutrino energy is exactly fixed. We have also taken into account here 
that in real experiments the size of the source and detector are much smaller 
than the smallest of the oscillation lengths $L^{\rm osc}_{jk}$, so that the 
localization condition \eqref{eq:local} is satisfied.

\section{\MB\ neutrinos in the intermediate wave packet formalism}
\label{sec:QM}

Although Eq.~\eqref{eq:infinite-wp-P1} shows that neutrino oscillations are not
inhibited by the energy constraints implied by the \MB\ effect, the assumption
of an exactly fixed neutrino energy is certainly unrealistic. Therefore, we
will now proceed to a more accurate treatment of \MB\ neutrinos using an
intermediate wave packet model~\cite{Giunti:1991ca,Giunti:1991sx,Kiers:1995zj,
Giunti:1997wq,Giunti:2003ax}. In this approach, the propagating neutrino is 
described by a superposition of mass eigenstates, each of which is in turn a 
wave packet with a finite momentum width. With the assumption of Gaussian wave 
packets, Giunti, Kim and Lee~\cite{Giunti:1991sx,Giunti:1997wq} obtain the
following expression for the $\bar{\nu}_e$ survival probability in the
approximation of ultra-relativistic neutrinos:
\begin{align}
  P_{ee}
    &= \sum_{j,k} |U_{ej}|^2 |U_{ek}|^2
       \exp\bigg[
             - 2\pi i \frac{L}{L^{\rm osc}_{jk}}
             - \bigg( \frac{L}{L^{\rm coh}_{jk}} \bigg)^2
             - 2\pi^2 \xi^2 \bigg( \frac{1}{2 \sigma_p L^{\rm osc}_{jk}} \bigg)^2
           \bigg].
  \label{eq:QM-P1}
\end{align}
Here
\begin{align}
  L^{\rm coh}_{jk} &= \frac{2 \sqrt{2} E^2}{\sigma_p |\Delta m_{jk}^2|} 
  \label{eq:Lcoh}
\end{align}
are the partial coherence lengths, $\sigma_p$ being the effective momentum 
uncertainty of the neutrino state, and the oscillation lengths 
$L^{\rm osc}_{jk}$ are given by Eq.~\eqref{eq:Losc}. $E$ is the energy 
that a massless neutrino emitted in the same process would have, and the 
$\mathcal{O}(1)$ parameter $\xi$ quantifies the deviation of the actual 
energies of massive neutrinos from this value. Since the energy uncertainty 
is very small for \MB\ neutrinos, the mass eigenstates differ in momentum, 
but hardly in energy, so that $\xi$ should be negligibly small in our case.

One can see that the first term in the exponent of Eq.~\eqref{eq:QM-P1}
is the standard oscillation phase. The second term yields a decoherence 
factor, which describes the suppression of oscillations due to the wave 
packet separation. For conventional neutrino experiments with non-negligible 
$\xi$, the third term implements a localization condition by suppressing 
oscillations if the spatial width $\sigma_x = 1 / 2\sigma_p$ of the neutrino 
wave packet is much larger than the oscillation length $L^{\rm osc}_{jk}$ 
(cf. Eqs.~(\ref{eq:Losc}) and (\ref{eq:local})). However, we have seen that, 
due to the smallness of $\xi$, the intermediate wave packet formalism predicts 
this condition to be irrelevant for oscillations of \MB\ neutrinos.

\section{\MB\ neutrinos in the external wave packet formalism}
\label{sec:QFT}

\begin{figure}
  \begin{center}
    \includegraphics{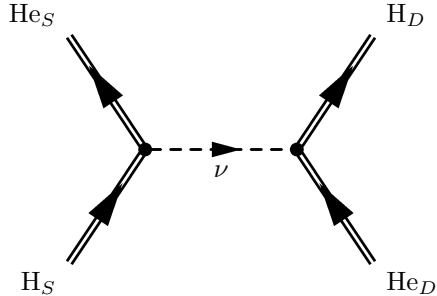}
  \end{center}
  \caption{Feynman diagram for neutrino emission and absorption
           in the $^3$H\,--$^3$He system.}
  \label{fig:feyn}
\end{figure}

In the derivation of the quantum mechanical result discussed in the previous 
section, certain assumptions had to be made on the properties of the neutrino 
wave packets, in particular on the parameters $\sigma_p$ and $\xi$. We will 
now proceed to the discussion of a QFT  
approach~\cite{Jacob:1961,Sachs:1963,Giunti:1993se,Rich:1993wu,Grimus:1996av,
Grimus:1998uh,Grimus:1999ra,Cardall:1999ze,Beuthe:2001rc,Beuthe:2002ej}, in 
which these quantities will be automatically determined from the properties of 
the source and the detector. 

Our calculation will be based on the Feynman diagram shown in 
Fig.~\ref{fig:feyn}, in which the neutrino is described as an internal line. 
We take the external particles to be confined by quantum mechanical harmonic
oscillator potentials to reflect the fact that they are bound in a crystal
lattice. Typical values for the harmonic oscillator frequencies are of the
order of the Debye temperature $\Theta_D \sim 600\ \mathrm{K}\simeq 0.05$ eV 
of the respective crystals~\cite{Raghavan:2006xf,Potzel:2006ad}. Although this 
simplistic treatment neglects the detailed structure of the solid state 
lattice, it is known to correctly reproduce the main features of the 
conventional \MB\ effect~\cite{Lipkin:1973}, and since we are interested 
mainly in the oscillation physics and not in the exact overall process rate, 
it is sufficient for our purposes. As only recoil-free neutrino emission 
and absorption are of interest to us, we can neglect thermal excitations and 
consider the parent and daughter nuclei in the source and detector to be in 
the ground states of their respective harmonic oscillator potentials. 

In Sec.~\ref{sec:QFT-minimal}, we will develop our formalism and derive
an expression for the rate of the combined process of \MB\ neutrino emission, 
propagation and absorption. In 
\mbox{Secs.~\ref{sec:QFT-inhom} -- \ref{sec:QFT-nat}} we will then discuss 
in detail the effects of different line broadening mechanisms.

\subsection{The formalism}
\label{sec:QFT-minimal}

Let us denote the harmonic oscillator frequencies for tritium and helium in
the source by $\omega_{\H,S}$ and $\omega_{\He,S}$ and those in the detector
by $\omega_{\H,D}$, and $\omega_{\He,D}$. In general, these are four different
numbers because $^3\H$ and $^3\He$ have different chemical properties, and
because their different abundances in the source and detector imply 
$\omega_{\H,S} \neq \omega_{\H,D}$ and $\omega_{\He,S} \neq \omega_{\He,D}$.
We ignore possible anisotropies of the oscillator frequencies because their 
inclusion would merely lengthen our formulas without giving new insights into
the oscillation phenomenology. The normalized wave functions of the ground 
states of the three-dimensional harmonic oscillators $\ket{\psi_{A,B,0}}$ are
given by
\begin{align}
  \psi_{A,B,0}(\vec{x}, t) = \bigg[\frac{m_A \omega_{A,B}}{\pi}\bigg]^\frac{3}{4}
  \exp\bigg[\! -\frac{1}{2} m_A \omega_{A,B} |\vec{x} - \vec{x}_B|^2 \bigg] \cdot e^{-i E_{A,B} t},
  \label{eq:HO-WF-gs}
\end{align}
where $A = \{ \H, \He \}$ distinguishes the two types of atoms and $B =
\{ S, D \}$ distinguishes between quantities related to the source and to the
detector. The masses of the tritium and $^3\He$ atoms are denoted by $m_\H$
and $m_\He$, and the coordinates of the lattice sites at which the atoms are
localized in the source and in the detector are $\vec{x}_S$ and $\vec{x}_D$.
The energies $E_{A,B}$ of the external particles are not exactly fixed due to
the line broadening mechanisms discussed in Sec.~\ref{sec:qualitative}, but
follow narrow distribution functions, which are centered around $E_{A,B,0} =
m_A + \frac{1}{2} \omega_{A,B}$. For the differences of these mean 
energies of tritium and helium atoms in the source and detector we will 
use the notation 
\begin{align}  
E_{S,0} = E_{\H,S,0} - E_{\He,S,0}\,,\qquad\qquad 
E_{D,0} = E_{\H,D,0} - E_{\He,D,0}\,.  
\label{eq:diff1}
\end{align}  

Before proceeding to calculate the overall rate of the process of neutrino 
production, propagation and detection, we compute the expected rates of 
the \MB\ neutrino production and detection treated as separate processes,  
ignoring neutrino oscillations. This calculation is very instructive, and
we will use its result as a benchmark for comparison with our subsequent 
QFT calculations. 

The effective weak interaction Hamiltonians for the neutrino production 
and detection $H_S^+$ and $H_D^-$ are given by Eqs. (\ref{eq:Hs+}) and 
(\ref{eq:Hd-}) of appendix C. We will first assume that the neutrino emitted 
in the recoil-free production process (\ref{eq:prod}) is monochromatic,
i.e.~neglect the natural linewidth as well as all broadening effects. Likewise,
we will neglect now the absorption line broadening effects in the recoilless 
detection process (\ref{eq:abs}). A straightforward calculation gives 
for the rate of recoilless neutrino production 
\begin{align}
\Gamma_p = \Gamma_0\,X_S\,,
\label{eq:Gammap}
\end{align}
where 
\begin{align}
\Gamma_0 = \frac{G_F^2 \cos^2\theta_c}{\pi}\; |\psi_e(R)|^2 \,m_e^2\, 
\left(|M_V|^2+g_A^2 |M_A|^2\right)\,\left(\frac{E_{S,0}}{m_e}\right)^2
\kappa_S 
\label{eq:Gamma0}
\end{align}
with $G_F$ the Fermi constant, $\theta_c$ the Cabibbo angle, $m_e$ 
the electron mass,  $M_V$ and $M_A$ the vector and axial-vector (or Fermi 
and Gamow-Teller) nuclear matrix elements and $g_A\simeq 1.25$ the 
axial-vector coupling constant. Note that for the allowed beta transitions in 
the $^3$H\,--$^3$He system, $M_V=1$ and $M_A\approx \sqrt{3}$. The quantity 
$\psi_e(R)$ is the value of the anti-symmetrized atomic wave function of
$^3\He$ at the surface of the nucleus. The factor $\kappa_S$ takes into
account that the spectator electron which is initially in the $1s$ atomic
state of $^3\H$ ends up in the $1s$ state of $^3\He$. It is given by 
the overlap integral of the corresponding atomic wave functions:
\begin{align}
\kappa_S~=~\Big|\int \Psi_{Z=2,S}(\vec{r})^* \,\Psi_{Z=1,S}(\vec{r})\, d^3 r
\,\Big|^2\,.
\label{eq:kappaS}
\end{align}
The factor $X_S$ in Eq.~(\ref{eq:Gammap}) is defined as 
\begin{align} 
X_S = 8\left(\eta_S+\frac{1}{\eta_S}\right)^{-3}
e^{-\frac{p^2}{\sigma_{pS}^2}}\,\equiv\,Y_{S}\,e^{-\frac{p^2}{\sigma_{pS}^2}}\,,
\label{eq:Xs}
\end{align} 
where $p=\sqrt{E_{S,0}^2-m^2}$ is the neutrino momentum%
\footnote{Since in this calculation we ignore neutrino oscillations, we 
also  neglect the neutrino mass differences.}, 
and 
\begin{align} 
\eta_S = \sqrt{\frac{m_\H \,\omega_{\H,S}}{m_\He\, \omega_{\He,S}}}\,,
\qquad\qquad \sigma_{pS}^2 = m_\H \,\omega_{\H,S}+m_\He\, \omega_{\He,S}\,.
\label{eq:etaS}
\end{align} 
The energy spectrum $\rho(E)$ of the emitted \MB\ neutrinos in the 
considered approximation is 
\begin{align} 
\rho(E) = \Gamma_0 \, X_S \, \delta(E-E_{S,0})\,.
\label{eq:rhoE}
\end{align}

For the cross section of the recoilless detection process (\ref{eq:abs}) 
we obtain 
\begin{align}
\sigma(E) = B_0\,X_D\,\delta(E-E_{D,0})\,,
\label{eq:sigma}
\end{align}
where 
\begin{align}
B_0=4\pi G_F^2 \cos^2\theta_c\; |\psi_e(R)|^2  
\left(|M_V|^2+g_A^2 |M_A|^2\right)\,\kappa_D\,. 
\label{eq:B0}
\end{align}
The factor $\kappa_D$ here is defined similarly to $\kappa_S$ in 
Eq.~\eqref{eq:kappaS}. Note that in the approximation of hydrogen-like 
atomic wave functions one has 
$\kappa_S=\kappa_D=512/729\simeq 0.7$. 
The factor $X_D$ in Eq.~\eqref{eq:sigma} is defined similarly to the 
corresponding factor for the 
production process, i.e.  
\begin{align} 
X_D = 8\left(\eta_D+\frac{1}{\eta_D}\right)^{-3}
e^{-\frac{p^2}{\sigma_{pD}^2}}\,\equiv\,Y_{D}\, e^{-\frac{p^2}{\sigma_{pD}^2}} 
\label{eq:Xd}
\end{align} 
with 
\begin{align} 
\eta_D = \sqrt{\frac{m_\H \,\omega_{\H,D}}{m_\He\, \omega_{\He,D}}}\,,
\qquad\qquad \sigma_{pD}^2 = m_\H \,\omega_{\H,D}+m_\He\, \omega_{\He,D}\,.
\label{eq:etaD}
\end{align} 
The \MB\ neutrino production rate $\Gamma_p$ and detection cross section 
$\sigma(E)$ differ from those previously obtained for unbound parent and 
daughter nuclei respectively in Refs.~\cite{Bahcall:1961} and 
\cite{Mikaelyan:1967} by the factors $X_S$ and $X_D$. Note that in the 
limit $m_\H \,\omega_{\H,S}=m_\He\, \omega_{\He,S}$, $m_\H \,\omega_{\H,D}=
m_\He\, \omega_{\He,D}$, the pre-exponential factors $Y_S$ and $Y_D$ in 
Eqs.~(\ref{eq:Xs}) and (\ref{eq:Xd}) become equal to unity, so that $X_S$ and 
$X_D$ reduce to the exponentials, which are merely the recoil-free fractions 
in the production and detection processes (see the discussion below).  

For unpolarized tritium nuclei in the source the produced neutrino flux is 
isotropic; therefore the spectral density of the neutrino flux at the 
detector located at a distance $L$ from the source is $\rho(E)/(4\pi L^2)$. 
The detection rate is thus
\begin{align}
\Gamma\,=\,\frac{1}{4\pi L^2}\int_0^\infty \rho(E)\sigma(E)\,dE \,=\, 
\frac{\Gamma_0\, B_0}{4\pi L^2}\,X_S X_D \;\delta(E_{S,0}-E_{D,0})\,.
\label{eq:rate1}
\end{align}
We see that it is infinite when the \MB\ resonance condition $E_{S,0}=E_{D,0}$ 
is exactly satisfied and zero otherwise, which is a consequence of our 
assumption of infinitely sharp emission and absorption lines. This assumption 
is certainly unphysical, and a realistic calculation should take into account 
the finite linewidth effects. We do that here by assuming Lorentzian energy 
distributions for the production and detection processes, which will be 
useful for comparison with the results of our subsequent QFT approach. In this 
approximation Eqs.~(\ref{eq:rhoE}) and (\ref{eq:sigma}) have to be replaced by 
\begin{align} 
\rho(E) = \Gamma_0 \,X_S\, 
\frac{\gamma_S/2\pi}{(E-E_{S,0})^2+\gamma_S^2/4}\,,
\qquad \sigma(E) = 
B_0\,X_D\,\frac{\gamma_D/2\pi}{(E-E_{D,0})^2+\gamma_D^2/4}\,,
\label{eq:rhoEsigma}
\end{align}
where $\gamma_S$ and $\gamma_D$ are the energy widths associated with 
production and detection. The combined rate of the neutrino production, 
propagation and detection process is then
\begin{align}
\Gamma\,= \,\frac{1}{4\pi L^2}\int_0^\infty \rho(E)\sigma(E)\,dE \,\simeq\, 
\frac{\Gamma_0\, B_0}{4\pi L^2}\,X_S X_D\;
\frac{(\gamma_S+\gamma_D)/2\pi}{(E_{S,0}-E_{D,0})^2+(\gamma_S+\gamma_D)^2/4}\,.
\label{eq:rate2}
\end{align}
As can be seen from this formula, the \MB\ resonance condition is 
\begin{align}
(E_{S,0}-E_{D,0})^2\ll(\gamma_S+\gamma_D)^2/4\,.
\label{eq:}
\end{align}
If it is satisfied, the neutrino detection cross section is enhanced by a 
factor of order $(\alpha Z m_e)^3/[p_e E_e(\gamma_S+\gamma_D)]$ compared to 
cross sections of non-resonant capture reactions $\bar{\nu}_e+A\to A'+e^+$ 
for neutrinos of the same energy (assuming the recoil-free fraction 
to be of order 1). For $\gamma_S+\gamma_D\sim 10^{-11}$ eV the enhancement 
factor can be as large as $10^{12}$. 

We now turn to the QFT treatment of the overall neutrino production,
propagation and detection process, first neglecting the line broadening
effects. We derive the corresponding transition amplitude from the matrix
elements of the weak currents in the standard way by employing the
coordinate-space Feynman rules to the diagram in Fig.~\ref{fig:feyn}.
For the external tritium and helium nuclei, we use the bound state wave
function $\psi_{A,B,0}(\vec{x}, t)$ from Eq.~\eqref{eq:HO-WF-gs}. We
obtain 
\begin{align}
  i \mathcal{A} &=
      \int\! d^3x_1 \, dt_1 \int\! d^3 x_2 \, dt_2 \,
      \bigg( \frac{m_\H  \omega_{\H, S}}{\pi} \bigg)^{\frac{3}{4}}
      \exp\bigg[ -\frac{1}{2} m_\H \omega_{\H,S}
      |\vec{x}_1 - \vec{x}_S|^2 \bigg]  \,  e^{-i E_{\H,S} t_1}  \nonumber\\
  &\hspace{1cm} \cdot
      \bigg( \frac{m_\He \omega_{\He,S}}{\pi} \bigg)^{\frac{3}{4}}
      \exp\bigg[ -\frac{1}{2} m_\He \omega_{\He,S}
      |\vec{x}_1 - \vec{x}_S|^2 \bigg]  \,  e^{+i E_{\He,S} t_1} \nonumber\\
  &\hspace{1cm} \cdot
      \bigg( \frac{m_\He \omega_{\He, D}}{\pi} \bigg)^{\frac{3}{4}}
      \exp\bigg[ -\frac{1}{2} m_\He \omega_{\He,D}
      |\vec{x}_2 - \vec{x}_D|^2 \bigg]  \,  e^{-i E_{\He,D} t_2} \nonumber\\
  &\hspace{1cm} \cdot
      \bigg( \frac{m_\H \omega_{\H,D}}{\pi} \bigg)^{\frac{3}{4}}
      \exp\bigg[ -\frac{1}{2} m_\H \omega_{\H,D}
      |\vec{x}_2 - \vec{x}_D|^2 \bigg]  \,  e^{+i E_{\H,D} t_2}  \nonumber\\
  &\hspace{1cm} \cdot \sum_j
      \mathcal{M}_S^\mu \mathcal{M}_D^{\nu *} |U_{ej}|^2 \, \int \! 
      \frac{d^4p}{(2\pi)^4}
      e^{-i p_0 (t_2 - t_1) + i \vec{p} (\vec{x}_2 - \vec{x}_1)} \nonumber\\
  &\hspace{1cm} \cdot
      \bar{u}_{e,S} \gamma_\mu (1 - \gamma^5) \,
      \frac{i (\slashed{p} + m_j)}{p_0^2 - \vec{p}^2 - m_j^2 + i\epsilon} \,
      (1 + \gamma^5) \gamma_\nu u_{e,D}.
  \label{eq:QFT-A1}
\end{align} 
The Dirac spinors for the external particles are denoted by $u_{A,B}$ 
with $A = \{ e, \H, \He \}$ and $B = \{ S, D \}$. Note that all spinors 
are non-relativistic, so that we can neglect their momentum dependence. 
The matrix elements $\mathcal{M}^\mu_S$ and $\mathcal{M}^\mu_D$ encode
the information on the bound state tritium beta decay and also on the
inverse process, the induced orbital electron capture which takes place
in the detector. They are given by
\begin{align}
  \mathcal{M}_{S,D}^\mu &= \frac{G_F  \cos\theta_c}{\sqrt{2}}  \, \psi_e(R) \,
                     \bar{u}_\He (M_V\, \delta^\mu_0 - g_A M_A \sigma_i \,
  \delta^{\mu}_i/\sqrt{3} ) u_\H\,\kappa_{S,D}^{1/2}\,.
  \label{eq:Mj}
\end{align}

The integrations over $t_1$ and $t_2$ in Eq.~\eqref{eq:QFT-A1} yield
energy-conserving $\delta$-functions at the neutrino production and detection 
vertices. The spatial integrals are Gaussian and can be evaluated after making 
the transformations $\vec{x}_1 \rightarrow \vec{x}_1 + \vec{x}_S$ and 
$\vec{x}_2 \rightarrow \vec{x}_2 + \vec{x}_D$. We obtain
\begin{align}
  i \mathcal{A} &= \mathcal{N} \int \! \frac{d^4p}{(2\pi)^4} \,
     2\pi \delta(p_0 - E_S) \, 2\pi \delta(p_0 - E_D) \,
     \exp\bigg[ -\frac{\vec{p}^2}{2 \sigma_p^2} \bigg] \nonumber\\
  &\hspace{1cm} \cdot
     \sum_j \mathcal{M}_S^\mu \mathcal{M}_D^{\nu *} |U_{ej}|^2
     \bar{u}_{e,S} \gamma_\mu (1 - \gamma^5) \,
     \frac{i (\slashed{p} + m_j) e^{i \vec{p} \vec{L}}}{p_0^2 - \vec{p}^2 - 
      m_j^2 + i\epsilon} \,
     (1 + \gamma^5) \gamma_\nu u_{e,D},
  \label{eq:QFT-A2}
\end{align}
where we have used the notation 
\begin{align}
E_S = E_{\H,S} - E_{\He,S}\,,\qquad\qquad
E_D = E_{\H,D} - E_{\He,D}\,,
\label{eq:EsEd}
\end{align} 
and introduced the baseline vector $\vec{L} = \vec{x}_D - \vec{x}_S$. The 
quantity $\sigma_p$, which is given by
\begin{align}
  \frac{1}{\sigma_p^2} =
    \frac{1}{m_\H \omega_{\H,S} + m_\He \omega_{\He,S}}
  + \frac{1}{m_\H \omega_{\H,D} + m_\He \omega_{\He,D}}\,,
  \label{eq:sigma-p}
\end{align}
can be interpreted as an effective momentum uncertainty of the neutrino.
Note that $\sigma_p^{-2}=\sigma_{pS}^{-2}+\sigma_{pD}^{-2}$. 
We have also defined a constant
\begin{align}
  \mathcal{N} &=
    \bigg( \frac{m_\H  \omega_{\H, S}}{\pi} \bigg)^{\frac{3}{4}}
    \bigg( \frac{m_\He \omega_{\He,S}}{\pi} \bigg)^{\frac{3}{4}}
    \bigg( \frac{m_\He \omega_{\He,D}}{\pi} \bigg)^{\frac{3}{4}}
    \bigg( \frac{m_\H  \omega_{\H, D}}{\pi} \bigg)^{\frac{3}{4}} \nonumber\\
  &\hspace{3cm} \cdot
    \bigg( \frac{2\pi}{m_\H \omega_{\H,S} + m_\He \omega_{\He,S}} 
                                            \bigg)^\frac{3}{2}
    \bigg( \frac{2\pi}{m_\H \omega_{\H,D} + m_\He \omega_{\He,D}} 
                                    \bigg)^\frac{3}{2},
\end{align}
containing the numerical factors from Eq.~\eqref{eq:HO-WF-gs} and coming from 
the integrals over $\vec{x}_1$ and $\vec{x}_2$. One of the $\delta$-functions 
in Eq.~\eqref{eq:QFT-A2} can now be used to perform the integration over 
$p_0$, thereby fixing $p_0$ at the value $p_0 = E_S = E_D$. To compute  
the remaining integral over the three-momentum $\vec{p}$, we
use a theorem by Grimus and Stockinger~\cite{Grimus:1996av}, which states the
following: Let $\psi(\vec{p})$ be a three times continuously differentiable
function on $\mathbb{R}^3$, such that $\psi$ itself and all its first and 
second derivatives decrease at least as $1/|\vec{p}|^2$ for $|\vec{p}| 
\rightarrow \infty$. Then, for any real number $A > 0$,
\begin{align}
  \int d^3p \, \frac{\psi(\vec{p}) \, e^{i \vec{p} \vec{L}}}{A - \vec{p}^2 + i\epsilon}
    \xrightarrow{|\vec{L}| \rightarrow \infty}
    -\frac{2 \pi^2}{L} \psi(\sqrt{A} \tfrac{\vec{L}}{L}) e^{i \sqrt{A} L}
    + \mathcal{O} (L^{-\frac{3}{2}}).
  \label{eq:Grimus}
\end{align}
The validity conditions are fulfilled in our case, so that in leading 
order in $1/L$ we have
\begin{align}
  i \mathcal{A} &= \frac{-i}{2L} \mathcal{N} \, \delta(E_S - E_D) \,
    \sum_j \exp\bigg[\! -\frac{E_S^2 - m_j^2}{2 \sigma_p^2} \bigg]
    \mathcal{M}_S^\mu \mathcal{M}_D^{\nu *} |U_{ej}|^2 \,
    e^{i \sqrt{E_S^2 - m_j^2} L}                        \nonumber\\
  &\hspace{6cm} \cdot
    \bar{u}_{e,S} \gamma_\mu (1 - \gamma^5) (\slashed{p}_j + m_j)
    (1 + \gamma^5) \gamma_\nu u_{e,D}\,,
  \label{eq:QFT-A3}
\end{align}
where the 4-vector $p_j$ is defined as $p_j = (E_S, (E_S^2 - m_j^2)^{1/2} \, 
\vec{L}/L)$. The Grimus-Stockinger theorem ensures that for $L\gg E_0^{-1}$, 
where $E_0$ is the characteristic neutrino energy, the intermediate-state 
neutrino is essentially on mass shell and its momentum points from the 
neutrino source to the detector.  

The transition probability $\mathcal{P}$ is obtained by summing 
$|\mathcal{A}|^2$ over the spins of the final states and averaging it over 
the initial-state spins. Note that no integration over final-state momenta 
is necessary because we consider transitions into discrete states.  The 
transition rate is obtained from $\mathcal{P}$ as $\Gamma = d\mathcal{P}/dT$, 
where $T$ is the total running time of the experiment. As we shall see, in 
the case of inhomogeneous line broadening $\mathcal{P}\propto T$ for large $T$, 
so that $\Gamma$ is independent of $T$ in that limit. The same is true for 
the homogeneous line broadening, except for the special case of the 
natural line width, for which the dependence on $T$ is more complicated 
(see Sec.~\ref{sec:QFT-nat}).

\subsection{Inhomogeneous line broadening}
\label{sec:QFT-inhom}

Inhomogeneous line broadening is due to stationary effects, such as 
impurities, lattice defects, variations in the lattice constant, etc. 
\cite{Potzel:2006ad,Balko:1997}. These effects are taken into 
account by summing the probabilities of the process for all possible 
energies of the external particles, weighted with the corresponding 
probabilities of these energies. In other words, one has to fold the 
probability or total rate of the process with the energy distributions of 
tritium and helium atoms in the source and detector, $\rho_{\He,S}(E_{\He,S})$, 
$\rho_{\H,D}(E_{\H,D})$, $\rho_{\H,S}(E_{\H,S})$ and $\rho_{\He,D}(E_{\He,D})$. 
We obtain
\begin{align}
\mathcal{P} &= 
\int_0^\infty
    \! dE_{\H,S} \, dE_{\He,S} \, dE_{\He,D} \, dE_{\H,D} \, \nonumber\\
  &\hspace{3cm} \cdot
    \rho_{\H,S}(E_{\H,S}) \, \rho_{\He,D}(E_{\He,D}) \,
    \rho_{\He,S}(E_{\He,S}) \, \rho_{\H,D}(E_{\H,D}) \,
    \overline{|\mathcal{A}|^2},
\end{align}
where $\overline{|\mathcal{A}|^2}$ is the squared modulus of the amplitude, 
averaged over initial spins and summed over final spins. Using the standard 
trace techniques to evaluate these spin sums and neglecting the momenta of 
the non-relativistic external particles, one finds 
\begin{align}
  \mathcal{P} 
    =& T\, \frac{G_F^4\,\cos^4\theta_c}{\pi L^2}\,|\psi_e(R)|^4 E_{S,0}^2\,
       (|M_V|^2 + g_A^2 |M_A|^2)^2\,Y_S Y_D \kappa_S \kappa_D
       \int_0^\infty \!\!\! dE_{\H,S} \, dE_{\He,S} \, dE_{\He,D} \, dE_{\H,D} \,
\nonumber\\
&\hspace{0.5cm} \cdot \delta(E_S - E_D)    
       \rho_{\H,S}(E_{\H,S}) \, \rho_{\He,D}(E_{\He,D}) \,
       \rho_{\He,S}(E_{\He,S}) \, \rho_{\H,D}(E_{\H,D})   \nonumber\\
    &\hspace{0.5cm} \cdot
       \sum_{j,k} |U_{ej}|^2 |U_{ek}|^2 \,
       \exp\bigg[\! -\frac{2 E_S^2 - m_j^2 - m_k^2}{2 \sigma_p^2} \bigg]
       e^{i \big(\sqrt{E_S^2 - m_j^2} - \sqrt{E_S^2 - m_k^2}\big) L}\,,
  \label{eq:QFT-Gamma1}
\end{align}
where $Y_S$ and $Y_D$ were defined in Eqs.~(\ref{eq:Xs}) and (\ref{eq:Xd}). 
Here we have taken into account that for $T \gg (E_S - E_D)^{-1}$ the 
squared $\delta$-function appearing in $\overline{|\mathcal{{A}}|^2}$ can 
be rewritten as%
\footnote{The expression $\delta(E_S-E_D)$ here should be understood as a 
$\delta$-like function of very small width. For $|E_S - E_D|\sim 10^{-11}$ 
eV, the condition $T \gg (E_S-E_D)^{-1}$ would require $T\gg 10^{-4}$ s, 
which should be very well satisfied in any realistic experiment. } 
\begin{align}
  [\delta(E_S - E_D)]^2
  \simeq \frac{1}{2\pi} \delta(E_S - E_D)
         \int_{-T/2}^{T/2} \! dt \, e^{i (E_S - E_D) t}
  =      \frac{T}{2\pi} \delta(E_S - E_D)\,.
  \label{eq:double-delta}
\end{align}
The overall process rate $\Gamma$ is then obtained from 
Eq.~(\ref{eq:QFT-Gamma1}) by simply dividing by $T$. Using the definitions 
of $\Gamma_0$ and $B_0$ given in Eqs.~(\ref{eq:Gamma0}) and (\ref{eq:B0}), 
one finds
\begin{align}
   \Gamma =& \frac{\Gamma_0 \,B_0}{4\pi L^2}\;Y_S Y_D  
       \int_0^\infty \! dE_{\H,S} \, dE_{\He,S} \, dE_{\He,D} \, dE_{\H,D} \,
\nonumber\\
&\hspace{0.5cm} \cdot \delta(E_S - E_D)    
       \rho_{\H,S}(E_{\H,S}) \, \rho_{\He,D}(E_{\He,D}) \,
       \rho_{\He,S}(E_{\He,S}) \, \rho_{\H,D}(E_{\H,D})   \nonumber\\
    &\hspace{0.5cm} \cdot
       \sum_{j,k} |U_{ej}|^2 |U_{ek}|^2 \,
       \exp\bigg[\! -\frac{2 E_S^2 - m_j^2 - m_k^2}{2 \sigma_p^2} \bigg]
       e^{i \big(\sqrt{E_S^2 - m_j^2} - \sqrt{E_S^2 - m_k^2}\big) L}.
  \label{eq:QFT-Gamma1a}
\end{align}

Before proceeding to the computation of the remaining integrations over the 
energy distributions of the external particles, 
let us discuss the expression in the last line in Eq.~\eqref{eq:QFT-Gamma1a}. 
In the approximation of ultra-relativistic (or nearly mass-degenerate) 
neutrinos, Eq.~\eqref{eq:relativistic-approx}, the last exponential becomes 
the standard oscillation phase factor $\exp(-2\pi i L / L^{\rm osc}_{jk})$ 
with the oscillation length defined in Eq.~\eqref{eq:Losc}. The additional
exponential suppression term $\exp[-(2 E_S^2 - m_j^2 - m_k^2)/2 \sigma_p^2]$
is an analogue of the well-known Lamb-\MB\ factor (or recoil-free
fraction)~\cite{Frauenfelder:1962,Lipkin:1973,Raghavan:2005gn}, which
describes the relative probability of recoil-free emission and absorption 
compared to the total emission and absorption probability. We see that  
for \MB\ neutrinos this factor depends not only on their energy, but also on 
their masses. Therefore, if two mass eigenstates, $\nu_j$ and $\nu_k$, do not
satisfy the relation $|\Delta m_{jk}^2| \ll \sigma_p^2$, the emission and 
absorption of the lighter mass eigenstate will be suppressed compared to 
the emission and absorption of the heavier one. This can be viewed as a 
reduced mixing of the two states, which in turn leads to a suppression 
of oscillations. To stress this point directly in our formulas, we rewrite
the corresponding factor as
\begin{align}
  \exp\bigg[ -\frac{(p^{\rm min}_{jk})^2}{\sigma_p^2} \bigg]
  \exp\bigg[ -\frac{|\Delta m_{jk}^2|}{2 \sigma_p^2} \bigg],
  \label{eq:Lamb-MB}
\end{align}
where $p^{\rm min}_{jk}$ is the smaller of the two momenta of the mass 
eigenstates $\nu_j$ and $\nu_k$,
\begin{align}
  (p^{\rm min}_{jk})^2 = E_S^2 - \max(m_j^2, m_k^2)\,.
\end{align}
The first exponential in Eq.~\eqref{eq:Lamb-MB} describes the 
suppression of the emission rate and the absorption cross section, i.e.~is 
a generalized Lamb-\MB\ factor, while the second one describes the 
suppression of oscillations. The condition $|\Delta m_{jk}^2| \lesssim 
2\sigma_p^2$ enforced by this second exponential can also be interpreted as 
a localization condition: Defining the spatial localization $\sigma_x \simeq 
1/2\sigma_p$, we can reformulate it as $L^{\rm osc}_{jk} \gtrsim 4\pi 
\sigma_x E_S/\sigma_p$. Since the generalized Lamb-\MB\ factor (the 
first factor in Eq.~(\ref{eq:Lamb-MB})) enforces $E_S \lesssim
\sigma_p$, this inequality is certainly fulfilled if $|L^{\rm osc}_{jk}|
\gtrsim 2\pi \sigma_x$ holds. The latter, stronger, localization condition 
is the one obtained in other external wave packet 
calculations~\cite{Giunti:1993se,Grimus:1998uh,Beuthe:2001rc} and is also 
equivalent to the one obtained in the intermediate wave packet 
picture~\cite{Giunti:1991sx,Giunti:1997wq} and discussed in Sec.~\ref{sec:QM}.

Let us now consider the integrations over the spectra of initial and final
states in Eq.~\eqref{eq:QFT-Gamma1a}. To evaluate these integrals, we need 
expressions for $\rho_{A,B}$, based on the physics of the inhomogeneous line 
broadening mechanisms. To a very good approximation, 
these effects cause a Lorentzian smearing of the energies of the external 
states~\cite{Potzel:PrivComm}, so that the energy distributions are 
\begin{align}
  \rho_{A,B}(E_{A,B}) &=
    \frac{\gamma_{A,B}/2\pi}{(E_{A,B} - E_{A,B,0})^2 + \gamma_{A,B}^2/4}\,,
  \label{eq:QFT-Lorentzian}
\end{align}
where, as before, $A = \{ \H, \He \}$, $B = \{ S, D \}$ and
$E_{A,B,0} = m_A + \frac{1}{2} \omega_{A,B}$. After evaluating the four
energy integrals in Eq.~\eqref{eq:QFT-Gamma1a} (see appendix
\ref{sec:appendix-inhom} for details), we obtain
\begin{align}
   \Gamma =& \frac{\Gamma_0 \,B_0}{4\pi L^2}\;Y_S Y_D \,\frac{1}{2\pi}\,
       \sum_{j,k} |U_{ej}|^2 |U_{ek}|^2 \,
       \exp\bigg[ -\frac{(p^{\rm min}_{jk})^2}{\sigma_p^2} \bigg]
    \exp\bigg[ -\frac{|\Delta m_{jk}^2|}{2 \sigma_p^2} \bigg]   \nonumber\\
    &\hspace{1cm} \cdot
 \frac{1}{E_{S,0} - E_{D,0} \pm i \,  \tfrac{\gamma_S - \gamma_D}{2}} \,
       \Bigg[
         \frac{\gamma_D A_{jk}^{(S)}}
           {E_{S,0} - E_{D,0} \pm i \, \tfrac{\gamma_S + 
\gamma_D}{2}}
       + \frac{\gamma_S A_{jk}^{(D)}}
           {E_{S,0} - E_{D,0} \mp i \,  \tfrac{\gamma_S + 
\gamma_D}{2}}
       \Bigg]\,.
  \label{eq:QFT-Gamma3}
\end{align}
In deriving this expression we have used the fact that the generalized 
Lamb-\MB\ factor is almost constant over the resonance region and can thus 
be approximated by its value at $\bar{E} = \frac{1}{2} (E_{S,0} + E_{D,0})$. 
The quantities $A_{jk}^{(B)}$ in Eq.~\eqref{eq:QFT-Gamma3} are given by
\begin{align}
  A_{jk}^{(B)}
    = \exp\bigg[ -i \frac{\Delta m_{jk}^2}{2(E_{B,0} \pm i \, 
    \tfrac{\gamma_B}{2})}\, L \bigg]  
&\simeq \exp\bigg[ -2\pi i \frac{L}{L^{\rm osc}_{B,jk}} \bigg] \, 
           \exp\bigg[ - \frac{L}{L^{\rm coh}_{B,jk}} \bigg]\,.
  \label{eq:inhom-A}
\end{align}
In Eqs.~(\ref{eq:QFT-Gamma3}) and (\ref{eq:inhom-A}) the upper (lower) 
signs correspond to $\Delta m_{jk}^2>0$ ($\Delta m_{jk}^2<0$). The oscillation 
and coherence lengths in (\ref{eq:inhom-A}) are defined in analogy with 
Eqs.~\eqref{eq:Losc} and \eqref{eq:Lcoh}:
\begin{align}
  L^{\rm osc}_{B,jk} = \frac{4\pi E_{B,0}}{\Delta m_{jk}^2}\simeq 
\frac{4\pi \bar{E}}{\Delta m_{jk}^2}\,,
\qquad\qquad
  L^{\rm coh}_{B,jk} = \frac{4 E_{B,0}^2}{\gamma_B |\Delta m_{jk}^2|} \simeq 
\frac{4 \bar{E}^2}{\gamma_B |\Delta m_{jk}^2|}\,, 
  \label{eq:Lcoh2}
\end{align}
We see that Eq.~\eqref{eq:QFT-Gamma3} depends not on the individual energies 
and widths of all external states separately, but only on the combinations 
$E_{B,0} = E_{\H,B,0} - E_{\He,B,0}$ and $\gamma_B=\gamma_{\H,B} + 
\gamma_{\He,B}$. In the limit of no neutrino oscillations, i.e.~when 
all $\Delta m_{jk}^2=0$ or $U_{aj}=\delta_{aj}$, Eq.~\eqref{eq:QFT-Gamma3} 
reproduces the no-oscillation result (\ref{eq:rate2}) obtained in our 
calculation of the \MB\ neutrino production and detection rates treated 
as separate processes. 

If the localization condition $|\Delta m_{jk}^2|\ll 2\sigma_p^2$ is satisfied 
for all $j$ and $k$, as it is expected to be the case in realistic experiments, 
one can pull the generalized Lamb-\MB\ factor out of the sum in  
Eq.~(\ref{eq:QFT-Gamma3}) and replace the localization exponentials by 
unity, which yields 
\begin{align}
   \Gamma \simeq  \frac{\Gamma_0 \,B_0}{4\pi L^2}\;Y_S Y_D \,
       \exp\bigg[ -\frac{E_{S,0}^2-m_0^2}{\sigma_p^2} \bigg]\,
 \sum_{j,k} |U_{ej}|^2 |U_{ek}|^2 \, I_{jk} \,.
  \label{eq:QFT-Gamma3a}
\end{align}
Here $m_0$ is an average neutrino mass and $I_{jk}$ is defined in 
Eq.~(\ref{eq:Ijk}). In realistic situations, it is often sufficient to 
consider two-flavour approximations to this expression. Indeed, at baselines 
$L \simeq 10$~m which 
are suitable to search for oscillations driven by $\theta_{13}$, the ``solar" 
mass squared difference $\Delta m_{21}^2$ is inessential, whereas for longer 
baselines around $L\simeq 300$~m, which could be used to study the oscillations 
driven by the parameters $\Delta m_{21}^2$ and $\theta_{12}$, the subdominant 
oscillations governed by $\Delta m_{31}^2$ and $\theta_{13}$ are in the 
averaging regime, leading to an effective 2-flavour oscillation probability. 
In both cases one therefore needs to evaluate 
\begin{align}
  &\sum_{j,k = 1, 2} |U_{ej}|^2 |U_{ek}|^2 \, I_{jk}
     =  \frac{(\gamma_S + \gamma_D) / 2\pi}
          {(E_{S,0} - E_{D,0})^2 + \frac{(\gamma_S + \gamma_D)^2}{4}}
     \Bigg\{
       (c^4 + s^4) + \frac{c^2 s^2}{2} \Big[ A^{(S)} + A^{(D)} + c.c. \Big]
     \Bigg\}                                    \nonumber\\
  &\hspace{0.5cm}
     - \frac{c^2 s^2 / 4\pi}{(E_{S,0} - E_{D,0})^2 + \frac{(\gamma_S + \gamma_D)^2}{4}}
       \Bigg[
         \frac{(A^{(S)} - A^{(D)})
           \big[
               (E_{S,0} - E_{D,0})(\gamma_S - \gamma_D)
             + i \frac{(\gamma_S + \gamma_D)^2}{2}
           \big]}
         { E_{S,0} - E_{D,0} + i \frac{\gamma_S - \gamma_D}{2}}
         + c.c.
       \Bigg],
  \label{eq:inhom-2f}
\end{align}
where $A^{(B)}$ ($B=S,D$) denotes the value of $A^{(B)}_{jk}$ corresponding to 
the appropriate fixed $\Delta m_{jk}^2\equiv \Delta m^2$ (which is defined 
here to be positive, i.e.~$\Delta m^2=|\Delta m_{31}|^2$ or $\Delta m_{21}^2$), 
$s = \sin\theta$ and $c = \cos\theta$, with $\theta$ being the relevant 
two-flavour mixing angle. 

As in the full three-flavour framework, in the absence of oscillations, 
i.e.~for $\Delta m^2 = 0$ or $\theta = 0$, Eqs.~\eqref{eq:QFT-Gamma3a} and 
(\ref{eq:inhom-2f}) reproduce the no-oscillation rate of Eq.~(\ref{eq:rate2}).
With oscillations included, the first line of Eq.~\eqref{eq:inhom-2f} 
factorizes into the Lorentzian times the $\bar{\nu}_e$ survival probability, 
which in general contains decoherence factors. Such a factorization does 
\emph{not} occur in the second line because the first term in the numerator 
in the square brackets is not proportional to $\gamma_S + \gamma_D$.
This term, containing a product of three small differences, is typically 
small compared to the other terms (at least when the \MB\ resonance 
condition $|E_{S,0} - E_{D,0}| \ll (\gamma_S + \gamma_D)/2$ is satisfied).
Still, it is interesting to observe that a naive factorization of $\Gamma$ 
into a no-oscillation transition rate and the $\bar{\nu}_e$ survival 
probability is not possible when this term is retained.

In all physically relevant situations, however, the whole second line of 
Eq.~\eqref{eq:inhom-2f} is negligible because so is $A^{(S)} - A^{(D)}$. 
Retaining only the contribution of the first line in Eq.~\eqref{eq:inhom-2f}, 
from Eq.~\eqref{eq:QFT-Gamma3a} one finds 
\begin{align}
   \Gamma \simeq &  \frac{\Gamma_0 \,B_0}{4\pi L^2}\;Y_S Y_D \,
       \exp\bigg[ -\frac{E_{S,0}^2-m_0^2}{\sigma_p^2} \bigg]\,
       \frac{(\gamma_S + \gamma_D) / 2\pi}
        {(E_{S,0} - E_{D,0})^2 + \frac{(\gamma_S + \gamma_D)^2}{4}} 
       \nonumber \\
& \hspace*{2cm} 
   \cdot \bigg\{1 - 2 s^2 c^2\bigg[1-\frac{1}{2}
( e^{-\alpha_S L} + e^{-\alpha_D L}) 
\cos\bigg(\frac{\Delta m^2 L}{4\bar{E}}\bigg)\bigg]\bigg\}\,,
  \label{eq:P-inhom-rel-2f-approx}
\end{align}
where $\alpha_{S,D} = (\Delta m^2/4\bar{E}^2) \gamma_{S,D}$, 
so that $\exp[-\alpha_{S,D} L]=\exp[-L/L^{\rm coh}_{S,D}]$ are the decoherence 
factors (cf. Eqs.~\eqref{eq:inhom-A} and \eqref{eq:Lcoh2}). For realistic 
experiments, one expects the oscillation phase $(\Delta m^2/4\bar{E})L$ to be 
of order unity, so that $\alpha_{S,D} L\sim \gamma_{S,D}/\bar{E}\sim 10^{-15}$, 
and decoherence effects are completely negligible. The second line in  
Eq.~\eqref{eq:P-inhom-rel-2f-approx} then yields the standard 2-flavour 
expression for the $\bar{\nu}_e$ survival probability. 

As we have already pointed out, the contribution of the second line in 
\eqref{eq:inhom-2f} to $\Gamma$ is of order $(e^{-\alpha_S L}-e^{-\alpha_D L})$ 
and therefore completely negligible. It is interesting to ask if there are any 
conceivable situations in which the decoherence exponentials in 
Eq.~\eqref{eq:P-inhom-rel-2f-approx} should be kept, while the contribution of 
the second line in \eqref{eq:inhom-2f} can still be neglected. Direct 
inspection of Eq.~\eqref{eq:inhom-2f} shows that this is the case when 
$|E_{S,0}-E_{D,0}| \lesssim |\gamma_S+\gamma_D|$ with 
$\alpha_{S,D} L\gtrsim 1$ and $|\alpha_S- \alpha_D|L\ll 1$.

\subsection{Homogeneous line broadening}
\label{sec:QFT-hom}
Homogeneous line broadening is caused by various electromagnetic relaxation 
effects, including interactions with fluctuating magnetic fields in the lattice 
\cite{Potzel:2006ad,Coussement:1992}. Unlike inhomogeneous broadening, it 
affects equally all the emitters (or absorbers) and therefore cannot be taken 
into account by  averaging the unperturbed transition probability over the
appropriate energy distributions of the participating particles, as we have
done in the previous subsection. Instead, one has to modify already the
expression for the amplitude. Since the homogeneous broadening effects
are stochastic, a proper averaging procedure, adequate to the broadening
mechanism, has then to be employed. For the conventional \MB\ effect with
long-lived nuclei, a number of models of homogeneous broadening was studied in 
\cite{Coussement:1992,Coussement:1992b,Odeurs:1995,Balko:1997,Odeurs:1997}. 
In all the considered cases the Lorentzian shape of the emission and 
absorption lines has been obtained. The same models can be used 
in the case of \MB\ neutrinos; one therefore expects that in most 
of the cases of homogeneous broadening the overall neutrino production -- 
propagation -- detection rate will also have the Lorentzian resonance 
form, i.e.~will essentially coincide in form with Eq.~\eqref{eq:QFT-Gamma3}, 
or with its simplified version in which the difference between $A_{jk}^{(S)}$ 
and $A_{jk}^{(D)}$ is neglected. A notable exception, which we consider next, 
is the homogeneous broadening due to the natural linewidth. As we shall see, 
this case is special because the time interval during which the source is 
produced is small compared with the tritium lifetime.

\subsection{Neutrino \MB\ effect dominated by the natural linewidth}
\label{sec:QFT-nat}

Although in a \MB\ neutrino experiment with a tritium source and a $^3\He$ 
absorber inhomogeneous broadening as well as homogeneous line broadening 
different from the natural linewidth are by far dominant, we will now consider 
also the case in which the emission and absorption linewidths are determined 
by the decay widths of the unstable nuclei. Even though it is not clear if 
such a situation can be realized experimentally, it is still very interesting 
for theoretical reasons.

To take the natural linewidth of tritium into account, we modify our 
expression for the amplitude, Eq.~\eqref{eq:QFT-A1}, by including exponential 
decay factors in the $^3$H wave functions. For the tritium in the source, 
this factor has the form $\exp(-\gamma t/2)$, describing a decay starting at 
$t = 0$, the time at which the experiment starts.%
\footnote{This is also supposed to be the time at which the number of $^3$H 
atoms in the source is known. It is assumed that the source is created in 
a time interval that is short compared to the tritium mean lifetime 
$\gamma^{-1}=17.81$ years.}
For the tritium which is produced in the detector, the decay factor
is $\exp(-\gamma (T - t_2)/2)$, where $t_2$ is the production time and $T$ 
is the time at which the number of produced $^3$H atoms is counted. Note that 
$\gamma$ here is the total decay width of tritium, not the partial width 
for bound state beta decay. Since we are taking into account the finite
lifetime of tritium, we also have to restrict the domain of all time
integrations in $\mathcal{A}$ to the interval $[0, T]$ instead of
$(-\infty, \infty)$. We thus have to compute 
\begin{align}
  i \mathcal{A} &=
      \int\! d^3x_1 \int_0^{T}\! dt_1 \int\! d^3x_2 \int_0^{T}\! dt_2 \,
      \bigg( \frac{m_\H  \omega_{\H, S}}{\pi} \bigg)^{\frac{3}{4}}
      \exp\bigg[ -\frac{1}{2} m_\H \omega_{\H,S}
      |\vec{x}_1 - \vec{x}_S|^2 \bigg]  \,
      e^{-i E_{\H,S,0} t_1 - \frac{1}{2} \gamma t_1}               
\nonumber\\
  &\hspace{1cm} \cdot
      \bigg( \frac{m_\He \omega_{\He,S}}{\pi} \bigg)^{\frac{3}{4}}
      \exp\bigg[ -\frac{1}{2} m_\He \omega_{\He,S}
      |\vec{x}_1 - \vec{x}_S|^2 \bigg]  \,  e^{+i E_{\He,S,0} t_1} 
\nonumber\\
  &\hspace{1cm} \cdot
      \bigg( \frac{m_\He \omega_{\He, D}}{\pi} \bigg)^{\frac{3}{4}}
      \exp\bigg[ -\frac{1}{2} m_\He \omega_{\He,D}
      |\vec{x}_2 - \vec{x}_D|^2 \bigg]  \,  e^{-i E_{\He,D,0} t_2} 
\nonumber\\
  &\hspace{1cm} \cdot
      \bigg( \frac{m_\H \omega_{\H,D}}{\pi} \bigg)^{\frac{3}{4}}
      \exp\bigg[ -\frac{1}{2} m_\H \omega_{\H,D}
      |\vec{x}_2 - \vec{x}_D|^2 \bigg]  \,
      e^{+i E_{\H,D,0} t_2 - \frac{1}{2} \gamma (T - t_2)}         
\nonumber\\
  &\hspace{1cm} \cdot \sum_j
      \mathcal{M}_S^\mu \mathcal{M}_D^{\nu *} |U_{ej}|^2 \, \int \! \frac{d^4p}{(2\pi)^4}
      e^{-i p_0 (t_2 - t_1) + i \vec{p} (\vec{x}_2 - \vec{x}_1)} \nonumber\\
  &\hspace{1cm} \cdot
      \bar{u}_{e,S} \gamma_\mu (1 - \gamma^5) \,
      \frac{i (\slashed{p} + m_j)}{p_0^2 - \vec{p}^2 - m_j^2 + i\epsilon} \,
      (1 + \gamma^5) \gamma_\nu u_{e,D}
  \label{eq:QFT-A5}
\end{align}
with the same notation as in Sec.~\ref{sec:QFT-minimal}. This form for 
$\mathcal{A}$ can also be derived in a more rigorous way using the 
Weisskopf-Wigner approximation~\cite{Weisskopf:1930au,Weisskopf:1930ps,
Grimus:1998uh,Cohen:QM2}, as shown in 
appendix~\ref{sec:appendix-WeisskopfWigner}. After 
a calculation similar to the one described in Sec.~\ref{sec:QFT-minimal}, we 
find for the total probability for finding a tritium atom at the lattice 
site $\vec{x}_D$ in the detector after a time $T$:
\begin{align}
  \mathcal{P}
   &= \frac{\Gamma_0 \,B_0}{4\pi L^2}\;Y_S Y_D \,\frac{2}{\pi}\, 
     \sum_{j,k} \theta(T_{jk}) \, |U_{ej}|^2 |U_{ek}|^2 \,  
                                                           \nonumber\\
  &\hspace{1cm} \cdot
     \exp\bigg[\! -\frac{(p^{\rm min}_{jk})^2}{\sigma_p^2} \bigg]
     \exp\bigg[\! -\frac{|\Delta m_{jk}^2|}{2 \sigma_p^2} \bigg]
     e^{i \big( \sqrt{\bar{E}^2 - m_j^2} - \sqrt{\bar{E}^2 - m_k^2} \big) L}
                                                           \nonumber\\
  &\hspace{1cm} \cdot
     e^{-\gamma T_{jk}} e^{- L / L^{\rm coh}_{jk}} \,
     \frac{\sin\big[ \frac{1}{2} (E_{S,0} - E_{D,0}) (T - \frac{L}{v_j}) 
     \big]\sin\big[ \frac{1}{2} (E_{S,0} - E_{D,0}) (T - \frac{L}{v_k}) 
     \big]} {(E_{S,0} - E_{D,0})^2}
  \label{eq:QFT-P2}
\end{align}
In the derivation, which is described in more detail in 
appendix~\ref{sec:appendix-nat}, we have neglected the energy dependence
of the generalized Lamb-\MB\ factor and of the spinorial terms, 
approximating them by their values at $\bar{E} = \frac{1}{2} (E_{S,0} 
+ E_{D,0})$. Furthermore, we have expanded the oscillation phase around this 
average energy. These approximations are justified by the observation that 
these quantities are almost constant over the resonance region.

In Eq.~\eqref{eq:QFT-P2} the quantity $v_j =(\bar{E}^2 - m_j^2)^{1/2}/\bar{E}$ 
denotes the group velocity of the $j$th neutrino mass eigenstate, and the 
generalized Lamb-\MB\ factor is parameterized in the by now familiar form 
with $p^{\rm min}_{jk} = \bar{E}^2 - \max(m_j^2, m_k^2)$. Moreover, we 
have defined the quantity 
\begin{align}
  T_{jk} = \min\bigg( T - \frac{L}{v_j}, \, T - \frac{L}{v_k} \bigg)\,,
\end{align}
which corresponds to the total running time of the experiment, minus the time 
of flight of the heavier of the two mass eigenstates $\nu_j$ and $\nu_k$. The 
appearance of the step-function factor $\theta(T_{jk})$ in 
Eq.~\eqref{eq:QFT-P2} is related to the finite neutrino time of flight 
between the source and the detector and to the fact that the interference 
between the $j$th and $k$th mass components leading to oscillations is 
only possible if both have already arrived at the detector. As in
Sec.~\ref{sec:QFT-inhom}, decoherence exponentials appear, containing the 
characteristic coherence lengths 
\begin{align}
  \frac{1}{L^{\rm coh}_{jk}} &= \gamma \, \bigg| \frac{1}{v_j} - \frac{1}{v_k} 
\bigg|\,.
\end{align}
In the approximation of ultra-relativistic (or nearly mass-degenerate)
neutrinos, this becomes
\begin{align}
  L^{\rm coh}_{jk} = \frac{4 \bar{E}^2}{\gamma |\Delta m_{jk}^2|}\,, 
\end{align}
and is thus analogous to Eqs.~\eqref{eq:Lcoh} and \eqref{eq:Lcoh2}. 

While the first two lines of Eq.~\eqref{eq:QFT-P2} contain the standard 
oscillation terms, the generalized Lamb-\MB\ factor and some numerical 
factors, the expression in the third line is unique to \MB\ neutrinos in the 
regime of natural linewidth dominance. To interpret this part of the 
probability, it is helpful to consider the approximation of massless 
neutrinos, which implies $v_j = 1$ for all $j$ and thus $L^{\rm coh}_{jk} = 
\infty$. If we neglect the time of flight $L/v_j$ compared to the total 
running time of the experiment $T$, we find that the probability is 
proportional to 
\begin{align}
  e^{-\gamma T} \, \frac{\sin^2 [(E_{S,0} - E_{D,0}) \frac{T}{2}]}{(E_{S,0} 
- E_{D,0})^2}.
  \label{eq:T-dependence1}
\end{align}
The factor $\exp(-\gamma t)$ accounts for the depletion of $^3\H$ in
the source and for the decay of the produced $^3\H$ in the detector.

It is easy to see that for $\gamma = 0$ and $T \rightarrow \infty$, 
Eq.~\eqref{eq:QFT-Gamma1a} is recovered, except for the omitted averaging 
over the energies of the initial and final state nuclei. In particular,
we see that in this limit, due to the emerging $\delta$-function, the \MB\ 
effect can only occur if the resonance energies $E_{S,0}$ and $E_{D,0}$ match 
exactly. For finite $T$, in contrast, the matching need not be exact because 
of the time-energy uncertainty relation, which permits a certain detuning, as 
long as $|E_{S,0} - E_{D,0}| \lesssim 1/T$. In the case of strong inequality 
$|E_{S,0} - E_{D,0}| \ll 1/T$, Eq.~\eqref{eq:T-dependence1} can be 
approximated by 
\begin{align}
  T^2\, e^{-\gamma T}/4\,.
  \label{eq:T-dependence2}
\end{align}
It is crucial to note that the allowed detuning of $E_{S,0}$ and $E_{D,0}$ 
does \emph{not} depend on $\gamma$, contrary to what one might expect. 
Instead, the \MB\ resonance condition requires this detuning to be small 
compared to the reciprocal of the overall observation time $T$. Therefore, the 
natural linewidth is not a fundamental limitation to the energy resolution of 
a \MB\ neutrino experiment. There is a well-known analogue to this in quantum 
optics~\cite{Meystre:1980}, called subnatural spectroscopy. Consider an 
experiment, in which an atom is instantaneously excited from its ground 
state into an unstable state $\ket{b}$ by a strong laser pulse at $t=0$. 
Moreover, the atom is continuously exposed to electromagnetic radiation 
with a photon energy $E$, which can eventually excite it further into another 
unstable state $\ket{a}$. If, after a time $\tau$, the number of atoms in 
state $\ket{a}$ is measured, it turns out that the result is proportional 
to $1/[(E - \Delta E)^2 + (\gamma_a - \gamma_b)^2/4]$ rather than to naively 
expected $1/[(E - \Delta E)^2 + (\gamma_a + \gamma_b)^2/4]$, where $\Delta E$ 
is the energy difference between the two states, and $\gamma_a$, $\gamma_b$ 
are their respective widths. In our case, the state $\ket{b}$ corresponds to a 
$^3\H$ atom in the source and a $^3\He$ atom in the detector, while $\ket{a}$ 
corresponds to a $^3\He$ atom in the source and a $^3\H$ atom in the detector.
The initial excitation of state $\ket{b}$ corresponds to producing the tritium 
source and starting the \MB\ neutrino experiment, and the transition from 
$\ket{b}$ to $\ket{a}$ corresponds to the production, propagation and 
absorption of a neutrino. Since the difference of decay widths 
$\gamma_a-\gamma_b$ vanishes for \MB\ neutrinos,%
\footnote{We assume that the tritium nuclei in the source and detector 
have the same mean lifetime.} 
we see that $\gamma$ does not have any impact on the 
achievable energy resolution, in accordance with Eq.~\eqref{eq:T-dependence1}. 
Note that this is only true because the source is produced at one specific 
point in time, namely $t=0$ (more generally, during a time interval that 
is short compared to the tritium lifetime). In a hypothetical experiment, 
in which tritium is continuously replenished in the source, an additional 
integration of $\mathcal{P}$ over the production time would be required, and 
this would yield proportionality to $1/[(E_{S,0} - E_{D,0})^2 + 
\gamma^2]$, in full analogy with the corresponding result in quantum 
optics~\cite{Meystre:1980}. 

The $T$-dependence of $\mathcal{P}$, as given by Eq.~\eqref{eq:T-dependence2}
can be understood already from a classical argument. If we denote the number
of $^3\H$ atoms in the source by $N_S$ and the corresponding number in 
the detector by $N_D$, the latter obeys the following differential equation:
\begin{align}
  \dot{N}_D = -\dot{N}_S N_{0} P_{ee} \frac{\sigma(T)}{4\pi L^2} - \gamma N_D\,.
  \label{eq:ND-ODE}
\end{align}
Here $P_{ee}$ is the $\bar{\nu}_e$ survival probability, $N_0$ is the number 
of $^3$He atoms in the detector, which we consider constant (this is justified 
if the number of $^3$H atoms produced in the detector is small compared to the 
initial number of $^3$He), and $\sigma(T)$ is the absorption 
cross section. It depends on $T$ because, due to the Heisenberg principle, 
the accuracy to which the resonance condition has to be fulfilled is  
given by $T^{-1}$. If we describe this limitation by assuming the emission and 
absorption lines to be Lorentzians of width $1/T$, we find that for 
$|E_{S,0}-E_{D,0}|\ll T^{-1}$ the overlap integral is proportional 
to $T$, so that we can write $\sigma = s_0\, T$ with $s_0$ a constant. Using 
furthermore the fact that $N_S = N_{S,0} \exp(-\gamma T)$, the 
solution of Eq.~\eqref{eq:ND-ODE} is found as 
\begin{align}
  N_D = \frac{N_{S,0} N_0 \gamma P_{ee} s_0}{8\pi L^2}
                \, T^2 e^{-\gamma T}.
\end{align}
This expression has precisely the $T$-dependence given by 
Eq.~\eqref{eq:T-dependence2}.

\section{Discussion}
\label{sec:discussion}

Let us now summarize our results. We have studied the properties of 
recoillessly emitted and absorbed neutrinos (\MB\ neutrinos) in a plane wave 
treatment (Sec.~\ref{sec:qualitative}), in a quantum mechanical wave packet 
approach (Sec.~\ref{sec:QM}) and in a full quantum field theoretical 
calculation (Sec.~\ref{sec:QFT}). The plane wave treatment corresponds to
the standard derivation based on the same energy approximation. We have pointed
out, that for \MB\ neutrinos this approximation is justifiable, even though for
conventional neutrino sources it is generally considered to be inconsistent.
The wave packet approach is an extension of the plane wave treatment, which
takes into account the small but non-zero energy and momentum spread of the
neutrino. Finally, the QFT calculation is superior to the other two, in
particular, because no prior assumptions about the energies and momenta
of the intermediate-state neutrinos have to be made. These properties are
automatically determined from the wave functions of the external particles
in the source and in the detector. For these wave functions we used well
established approximations that are known to be good in the theory of
the standard \MB\ effect.

In all three approaches that have been discussed, we have consistently arrived
at the prediction that \MB\ neutrinos will oscillate, in spite of their very
small  energy uncertainty. The plane wave result, Eq.~\eqref{eq:relativistic-approx}, 
is actually the standard textbook expression for the $\bar{\nu}_e$ survival 
probability, and Eqs.~\eqref{eq:QM-P1},
\eqref{eq:QFT-Gamma3} and \eqref{eq:QFT-P2} are extensions of this expression, 
containing, in particular, decoherence and localization factors. We have found 
that these factors cannot suppress oscillations under realistic experimental 
conditions, but are very interesting from the theoretical point of view. 

Let us now compare the results of different approaches. First, we observe that 
the decoherence exponents in our QFT calculations are linear in
$L/L^{\rm coh}$, while in the quantum mechanical result, Eq.~\eqref{eq:QM-P1},
the dependence is quadratic. This behaviour can be traced back to the fact
that Gaussian neutrino wave packets have been assumed in the quantum
mechanical computation, while in our QFT approach we have employed the
Lorentzian line shapes, which are more appropriate for describing \MB\
neutrinos. The linear dependence of the decoherence exponents on
$L/L^{\rm coh}$ in the case of the Lorentzian neutrino energy
distribution has been previously pointed out in~\cite{Grimus:1998uh}. 

Even more striking than the differing forms of the decoherence exponentials  
is the fact that a localization factor of the form $\exp[-|\Delta m_{jk}^2|
/2\sigma_p^2]$ is present in Eqs.~\eqref{eq:QFT-Gamma3} and \eqref{eq:QFT-P2}, 
while the localization exponentials disappear from Eq.~\eqref{eq:QM-P1} in the 
limit $\xi \to 0$ which is relevant for \MB\ neutrinos. This shows that the 
naive quantum mechanical wave packet approach does not capture all features of 
\MB\ neutrinos. In particular, it neglects the differences between the 
emission (and absorption) probabilities of different mass eigenstates, which 
effectively may lead to a suppression of neutrino mixing. In realistic 
experiments, however, this effect should be negligible.

Another interesting feature of the exponential factors implementing the 
localization condition in our QFT calculations is that the corresponding 
exponents are linear in $|\Delta m_{jk}^2|$, whereas the dependence is 
quadratic (for $\xi\ne 0)$ in the quantum-mechanical expression 
\eqref{eq:QM-P1}. This can be attributed to the fact that we consider the 
parent and daughter nuclei in the source and detector to be in bound states 
with zero mean momentum (but non-zero {\it rms} momentum). This is also 
the reason why the $\sigma_p$-dependence of the localization exponents in 
Eqs.~\eqref{eq:QFT-Gamma3} and \eqref{eq:QFT-P2} is different from that in 
the quantum mechanical approach (namely, they depend on $|\Delta m_{jk}^2|
/2\sigma_p^2$ rather than $|\Delta m_{jk}^2|/2p\sigma_p$): for the considered 
bound states, the {\it rms} momentum is $\bar{p}\sim \sigma_p$, so that 
$\bar{p}\sigma_p\sim \sigma_p^2$. 

One more point to notice is that while the same quantity, the momentum 
uncertainty \mbox{$\sigma_p$}, enters into the decoherence and localization 
factors in the quantum mechanical formula~\eqref{eq:QM-P1}, this is not the 
case in the QFT approach, where the localization factors depend on $\sigma_p$, 
whereas the decoherence exponentials are determined by the (much smaller)  
energy uncertainty. In the case of natural line broadening this energy 
uncertainty  is given by the $^3$H decay width $\gamma$, while in all the other 
cases it is given by the widths of the neutrino emission and 
absorption lines, which are determined by the homogeneous and inhomogeneous 
line broadening effects taking place in the source and detector.  
 
The QFT results of Eqs.~\eqref{eq:QFT-Gamma3} and \eqref{eq:QFT-P2} describe 
not only the oscillation physics, but also the production and detection 
processes. These results can thus also be used for an approximate prediction 
of the total event rate expected in a \MB\ neutrino experiment. 
Both expressions contain the Lamb-\MB\ factor (or recoil-free
fraction), which describes the relative probability of recoilless decay 
and absorption of neutrinos. Moreover, they contain factors that suppress  
the overall process rate $\Gamma$ unless the emission and absorption lines 
overlap sufficiently well. In the case of inhomogeneous line broadening 
(Sec.~\ref{sec:QFT-inhom}) as well as for homogeneous broadening different 
from the natural linewidth effect, this is a Lorentzian factor, the same as 
in the no-oscillation rate  \eqref{eq:rate2}. It suppresses the transition rate 
if the peak energies of the emission and absorption lines differ by more than 
the combined linewidth $\gamma_S+\gamma_D$. We have, however, found that the 
factorization of the total rate into the no-oscillation rate including the 
overlap factor and the oscillation probability is only approximate. For the 
hypothetical case of an experiment in which the neutrino energy uncertainty is 
dominated by the natural linewidth $\gamma$ (Sec.~\ref{sec:QFT-nat}), we have 
found that the overlap condition does not depend on $\gamma$, but is rather 
determined by the reciprocal of the overall duration of the experiment $T$.  
Although this result may seem counterintuitive at first sight, it has a 
well-known analogy in quantum optics~\cite{Meystre:1980} and is related to 
the fact that the initial unstable particles in the source are produced in a 
time interval much shorter than their lifetime.

Notice that the overlap factors contained in our QFT-based results for the 
neutrino \MB\ effect governed by the natural linewidth and by other line 
broadening mechanisms, Eqs.~\eqref{eq:QFT-P2} and \eqref{eq:QFT-Gamma3a},  
are two well-known limiting representations of the $\delta$-function, which 
yield the energy-conserving $\delta$-function $\delta(E_{S,0}-E_{D,0})$ in the 
limits $T\to \infty$ or $\gamma_S+\gamma_D\to 0$, respectively.%
\footnote{Eq.~\eqref{eq:QFT-P2} yields $T\delta(E_{S,0}-E_{D,0})$ because it 
describes a probability rather than a rate.} One can see that in these limits 
both expressions reproduce, if one sets the $\bar{\nu}_e$ survival probability 
$P_{ee}$ to unity, the no-oscillation rate~\eqref{eq:rate1} obtained in the 
infinitely sharp neutrino line limit by treating the \MB\ neutrino production 
and detection as separate processes. 
Our QFT results thus generalize the results of the standard calculations 
and allow a more accurate and consistent treatment of both the production 
-- detection rate and the oscillation probability of \MB\ neutrinos. 

To conclude, we have performed a quantum field theoretic calculation of 
the combined rate of the emission, propagation and detection of \MB\ neutrinos 
for the cases of inhomogeneous and homogeneous neutrino line broadening. 
In both cases we found that the decoherence and localization damping factors 
present in the combined rate will not play any role in realistic experimental 
settings and therefore will not prevent \MB\ neutrinos from oscillating.

\begin{acknowledgments}
It is a pleasure to thank F.~v.~Feilitzsch, H.~Kienert, J.~Litterst, W.~Potzel,
G.~Raffelt and V.~Rubakov for very fruitful discussions. This work was in part
supported by the Transregio Sonderforschungsbereich TR27 ``Neutrinos and Beyond''
der Deutschen Forschungsgemeinschaft. JK would like to acknowledge support from
the Studienstiftung des Deutschen Volkes.
\end{acknowledgments}

\appendix
\section{Derivation of the transition rate for inhomogeneous line broadening}
\label{sec:appendix-inhom}

In this appendix we give details of the steps leading from 
Eq.~\eqref{eq:QFT-Gamma1a} to Eq.~\eqref{eq:QFT-Gamma3} when the 
energy densities of the external states $\rho_{A,B}$ are chosen to 
have the Lorentzian form, Eq.~(\ref{eq:QFT-Lorentzian}). 

First, we notice that the Lorentzians are sharply peaked, with their widths 
much smaller than the peak energies, and therefore the energy integrals 
in~\eqref{eq:QFT-Gamma1a} get their main contributions from the narrow 
intervals around these peaks. Since the peak energies, as well as their 
differences $E_{\H,S,0}-E_{\He,S,0}$ and $E_{\H,S,0}-E_{\He,S,0}$ 
that determine the neutrino energies, are much larger than the 
neutrino masses, one can employ the expansion $\sqrt{E_S^2-m_{i}^2}\simeq E_S
-m_{i}^2/2E_S$ in the exponents. 

Next, we make use of the identity 
\begin{align}
  \int_{-\infty}^\infty dE_a \, dE_b \,&
     \frac{\gamma_a/2\pi}{(E_a - E_{a,0})^2 + \frac{\gamma_a^2}{4}} \,
     \frac{\gamma_b/2\pi}{(E_b - E_{b,0})^2 + \frac{\gamma_b^2}{4}}
     \, f(E_a - E_b)                                         \nonumber\\
  &= \int_{-\infty}^\infty d(E_a - E_b) \,
     \frac{(\gamma_a + \gamma_b)/2\pi}{\big[ (E_a - E_b) - (E_{a,0} - E_{b,0})
           \big]^2 + \frac{(\gamma_a + \gamma_b)^2}{4}} \, f(E_a - E_b)\,,
  \label{eq:eff-Lorentzian}
\end{align}
that holds for any function $f(E)$ for which the integrals in 
(\ref{eq:eff-Lorentzian}) exist. To apply this formula to 
Eq.~\eqref{eq:QFT-Gamma1a}, we have to extend the domain of the energy 
integrals from the physical region $[\max_j(m_j),\infty)$ to the whole real 
axis, $(-\infty,\infty)$. This is again possible because the Lorentzians 
$\rho_{A,B}(E_{A,B})$ have very narrow widths and therefore ensure that the 
unphysical contributions are strongly suppressed, the error introduced by 
the extension of the integration interval being of order 
$\gamma_{S(D)}/E_{S(D),0}\sim 10^{-15}$. We can thus use 
Eq.~\eqref{eq:eff-Lorentzian} to perform two of the four energy integrations 
in Eq.~\eqref{eq:QFT-Gamma1a}. Of the remaining two, one is trivial due to 
the factor $\delta(E_S - E_D)$, so that the expression for $\Gamma$ becomes
\begin{align}
  \Gamma = \frac{\Gamma_0 \,B_0}{4\pi L^2}\;Y_S Y_D 
       \int_{-\infty}^\infty \! dE \,
       \frac{\gamma_S/2\pi}{(E - E_{S,0})^2 + \gamma_S^2/4} \,
       \frac{\gamma_D/2\pi}{(E - E_{D,0})^2 + \gamma_D^2/4} \,  \nonumber\\
    \hspace{1cm} \cdot
       \sum_{j,k} |U_{ej}|^2 |U_{ek}|^2 \,
       \exp\bigg[\! -\frac{2 E^2 - m_j^2 - m_k^2}{2 \sigma_p^2} \bigg]
       e^{-i \frac{\Delta m_{jk}^2}{2 E}}\,.
  \label{eq:QFT-Gamma2a}
\end{align}

\begin{figure}
  \vspace{0.3cm}
  \begin{center}
    \includegraphics[width=8cm]{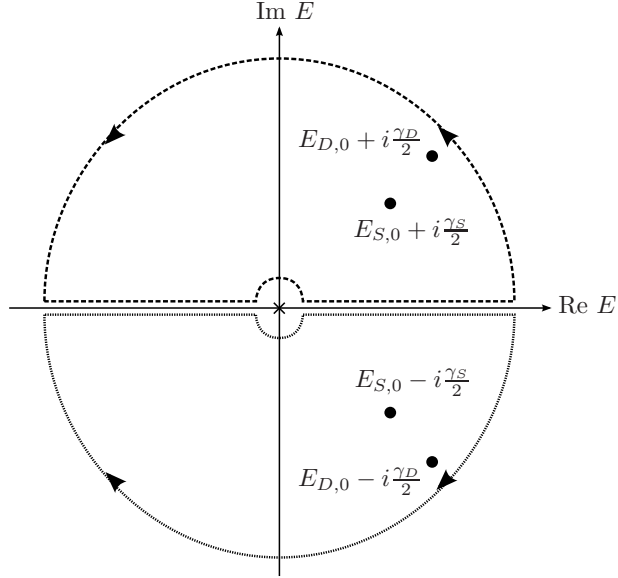}
  \end{center}
  \caption{Integration contours in the complex $E$ plane.}
  \label{fig:contours-3}
\end{figure}

Next, we pull the generalized Lamb-\MB\ factor out of the integral, 
replacing it by its value at $\bar{E} = (E_{S,0} + E_{D,0})/2$. This is 
justified by the observation that $\gamma_S, \gamma_D \sim 10^{-11}\
\text{eV} \ll \sigma_p \sim 10 \ \text{keV}$, so that the generalized 
Lamb-\MB\ factor is nearly constant over the region where the integrand 
is sizeable. We are thus left with the task to compute the expression 
\begin{align}
  I_{jk} \equiv \int_{-\infty}^\infty \! dE \,
  \frac{\gamma_S/2\pi}{(E - E_{S,0})^2 + \gamma_S^2/4} \,
  \frac{\gamma_D/2\pi}{(E - E_{D,0})^2 + \gamma_D^2/4} \,
       e^{-i \frac{\Delta m_{jk}^2}{2 E}}\,,
\label{eq:Ijk1}
\end{align}
which can be done by integration in the complex plane. 
The integrand of (\ref{eq:Ijk1}) has four poles, two above the real axis and 
two below, and an essential singularity at $E=0$ (see 
Fig.~\ref{fig:contours-3}). To circumvent the essential singularity, we choose 
the integration contour to consist of the real axis with a small interval 
$[-\varepsilon, \varepsilon]$ cut out, supplemented by a half-circle of radius 
$\varepsilon$ around the point $E=0$ and closed by a half-circle of large 
radius. The contribution of the small half-circle vanishes when its radius 
goes to zero provided that we avoid the point $E=0$ from above when 
$\Delta m_{jk}^2>0$ and from below when $\Delta m_{jk}^2<0$. Thus, we close 
the integration contour in the upper half-plane for $\Delta m_{jk}^2>0$ 
and in the lower half-plane for $\Delta m_{jk}^2<0$. 
The contribution from the large half-circle vanishes when its radius tends 
to infinity because the product of two Lorentzians goes to zero as $|E|^{-4}$ 
for $|E| \rightarrow \infty$, while the exponential becomes unity in this 
limit. Application of the residue theorem yields now 
\begin{align}
  I_{jk} &= \frac{1}{2\pi}
    \frac{1}{E_{S,0} - E_{D,0} \pm i \, \tfrac{\gamma_S - \gamma_D}{2}} 
\,
    \Bigg[
        \frac{\gamma_D A_{jk}^{(S)}}
             {E_{S,0} - E_{D,0} \pm i \,  \tfrac{\gamma_S + 
\gamma_D}{2}}
      + \frac{\gamma_S A_{jk}^{(D)}}
             {E_{S,0} - E_{D,0} \mp i \,  \tfrac{\gamma_S + 
\gamma_D}{2}}
    \Bigg],
  \label{eq:Ijk}
\end{align}
with the notation from Sec.~\ref{sec:QFT-inhom}. Here the upper (lower) signs  
correspond to $\Delta m_{jk}^2>0$ ($\Delta m_{jk}^2<0$). Inserting this result 
into Eq.~\eqref{eq:QFT-Gamma2a}, we obtain Eq.~\eqref{eq:QFT-Gamma3}.

\section{Derivation of the total transition probability for the case of
natural linewidth dominance}
\label{sec:appendix-nat}

In this appendix we describe the derivation leading from
Eq.~\eqref{eq:QFT-A5} to Eq.~\eqref{eq:QFT-P2}. The spatial integrals in 
Eq.~\eqref{eq:QFT-A5} are the same as those encountered in 
Sec.~\ref{sec:QFT-minimal} and yield the factor $\exp[-\vec{p}^2/2\sigma_p^2]
\exp[i \vec{p} \vec{L}]$. The time integrals can also be evaluated 
straightforwardly; however, unlike the corresponding integrals in 
Eq.~\eqref{eq:QFT-A1}, they do not give exact, but only approximate energy 
conserving factors for the production and detection processes. This behaviour 
can be ascribed to the non-zero width of the tritium states and the finite 
measurement time $T$. To evaluate the three-momentum integral over $\vec{p}$, 
we again employ the Grimus-Stockinger theorem and thus find 
\begin{align}
  i \mathcal{A}
  &= \frac{-i}{8 \pi^2 L} \mathcal{N} 
     \sum_j \mathcal{M}_S^\mu \mathcal{M}_D^{\nu *} |U_{ej}|^2 \int_
     {-\infty}^{\infty}\! dp_0 \,
     \bar{u}_{e,S} \gamma_\mu (1 - \gamma^5) (\slashed{p}_j + m_j)
     (1 + \gamma^5) \gamma_\nu u_{e,D}
                                                           \nonumber\\
  &\hspace{0.5cm} \cdot
     e^{-\frac{\gamma}{2} T} \,
     \frac{e^{-i (E_S - p_0) T - \frac{\gamma}{2} T } - 1}
          {p_0 - E_S + i \frac{\gamma}{2}} \,
     \frac{e^{ i (E_D - p_0) T + \frac{\gamma}{2} T } - 1}
          {p_0 - E_D + i \frac{\gamma}{2}} \,
     \exp\bigg[\! -\frac{p_0^2 - m_j^2}{2 \sigma_p^2} \bigg] \,
     e^{i \sqrt{p_0^2 - m_j^2} L},
  \label{eq:QFT-A6}
\end{align}
where the 4-vector $p_j$ is defined as $p_j = (p_0, (p_0^2 - m_j^2)^{1/2} \, 
\vec{L}/L)$. The exponential depending on $\sigma_p^2$, which will eventually 
lead to the generalized Lamb-\MB\ factor, can be approximated by its value 
at $\bar{E} = (E_S + E_D)/2$ because $\gamma \ll \sigma_p$ ensures that it is 
almost constant in the region from which the main contribution to the integral 
comes, namely the region where $|p_0 - E_S| \lesssim \gamma$ and $|p_0 - E_D| 
\lesssim \gamma$. The fact that this region is very narrow also allows us to 
pull the spinorial factors out of the integral and to expand the oscillation 
phase around $\bar{E}$: 
\begin{align}
  i \sqrt{p_0^2 - m_j^2} L \simeq
  i \sqrt{\bar{E}^2 - m_j^2} L
       + i \frac{L}{v_j} (p_0 - \bar{E})\,,
\end{align}
where $v_j = (\bar{E}^2 - m_j^2)^{1/2} / \bar{E}$. The integral over $p_0$ can 
then be evaluated by complex contour integration. The denominator has poles 
at $p_0 = E_S - i \gamma/2$ and $p_0 = E_D-i\gamma/2$, and the relevant terms 
in the numerator are
\begin{align}
  &\Big( e^{-i (E_S - p_0) T - \frac{1}{2} \gamma T} - 1 \Big)
   \Big( e^{i (E_D - p_0) T + \frac{1}{2} \gamma T} - 1 \Big) \,
   e^{i (p_0 - \bar{E}) \frac{L}{v_j}}
                                                                     \nonumber\\
  &\hspace{3cm}
    = \underbrace{ e^{i p_0 \frac{L}{v_j}}
          e^{-i (E_S - E_D) T - \bar{E} \frac{L}{v_j}} }_\textrm{(A)}
      \,-\, \underbrace{ e^{i p_0 (T + \frac{L}{v_j})}
   e^{-i E_S T - i \bar{E} \frac{L}{v_j} - \frac{1}{2} \gamma T} }_\textrm{(B)}
                                                                  \nonumber\\
  &\hspace{3cm} \phantom{=}
    \,-\, \underbrace{ e^{-i p_0 (T - \frac{L}{v_j})}
   e^{ i E_D T - i \bar{E} \frac{L}{v_j} + \frac{1}{2} \gamma T}}_\textrm{(C)}
 \,+\,\underbrace{ e^{i \frac{L}{v_j} p_0} e^{-i \bar{E} \frac{L}{v_j}} 
  }_\textrm{(D)}.
\end{align}
To close the integration contour, we add to the real axis a half-circle 
of infinite radius. For the terms labeled (A), (B) and (D), this half-circle 
has to lie in the upper half-plane, while for (C) it has to lie in the upper 
half-plane for $T < L/v_j$, and in the lower half-plane for $T > L/v_j$. 
As the integrand is holomorphic for ${\rm Im}(p_0) \geq 0$, only in this last 
case the integral can be non-zero. The residue theorem then yields 
\begin{align}
  i \mathcal{A}
  &= \frac{\mathcal{N}}{4 \pi L} \sum_j \theta(T - L/v_j) \,
     \mathcal{M}_S^\mu \mathcal{M}_D^{\nu *} |U_{ej}|^2 \cdot
     \exp\bigg[\! -\frac{\bar{E}^2 - m_j^2}{2 \sigma_p^2} \bigg] \,
     \bar{u}_{e,S} \gamma_\mu (1\!-\!\gamma^5) (\slashed{\bar{p}}_j + m_j)
     (1\!+\!\gamma^5)                          \nonumber\\
  &\hspace{0.2cm} 
     \cdot \gamma_\nu u_{e,D}
     e^{i \sqrt{\bar{E}^2 - m_j^2} L}
     \frac{e^{-\frac{1}{2} \gamma (T - \frac{L}{v_j})}
           e^{-\frac{i}{2} (E_S - E_D) T)}}{E_S - E_D}
     \bigg[ e^{-\frac{i}{2} (E_S - E_D) (T - \frac{L}{v_j})}
          - e^{\frac{i}{2} (E_S - E_D) (T - \frac{L}{v_j})} \bigg]\,,
  \label{eq:QFT-A7}
\end{align}
where now $\bar{p}_j = (\bar{E}, (\bar{E}^2 - m_j^2)^{1/2} \, \vec{L}/L)$, 
and $\theta(x)$ is the Heaviside step function. The total probability for 
finding a tritium atom at the lattice site $\vec{x}_D$ in the detector after 
a time $T$ is
\begin{align}
  \mathcal{P} = \overline{|\mathcal{A}|^2}\,,
  \label{eq:QFT-P1}
\end{align}
where the bar indicates the average over initial spins and the sum over final 
spins. Apart from these spin sums, no integration over the energy 
distributions of the initial and final state nuclei is necessary as long as 
only natural line broadening is taken into account, because we are dealing 
with transitions between discrete energy eigenstates. A straightforward 
evaluation of Eq.~\eqref{eq:QFT-P1} yields Eq.~\eqref{eq:QFT-P2}.

\section{Weisskopf-Wigner approach to the effects of the natural line 
width}
\label{sec:appendix-WeisskopfWigner}

In this appendix we use the Weisskopf-Wigner approach~\cite{Weisskopf:1930au,
Weisskopf:1930ps,Cohen:QM2,Grimus:1998uh} to derive Eq.~\eqref{eq:QFT-A5}, 
which has been the starting point for our discussion of \MB\ neutrinos in 
the regime of natural linewidth dominance. In particular, our aim is to 
substantiate the arguments dictating the form of the exponential decay 
factors by an explicit derivation.

We can write the Hamiltonian of the system as
$\Hamil = \Hamil_0 + e^{i \Hamil_0 t} \Hamil_1 e^{-i \Hamil_0 t}$,
where $\Hamil_1$ is the interaction-representation weak interaction 
Hamiltonian and $\Hamil_0$ is the remainder. In general, we will not
treat $H_1$ as a perturbation since we are ultimately interested in the
depletion of unstable states over time, which cannot be adequately described
in a perturbative approach. 

One can write an arbitrary state as $\ket{\psi(t)} = \sum_j c_j(t) 
\ket{\phi_j}$, where $\ket{\phi_j}$ are the eigenstates of $\Hamil_0$. 
The Schr\"odinger equation then gives the evolution equations for the 
coefficients $c_j(t)$: 
\begin{align}
  i \dot{c}_j(t) = \sum_k \bra{\phi_j} \Hamil_1 \ket{\phi_k} \, c_k(t)\,.
  \label{eq:cj-evolution}
\end{align}
For our purposes it will be convenient to slightly modify the notation and 
classify the different states according to their particle content,
as shown in Table~\ref{tab:states}. 
\begin{table}
  \begin{center}
    \parbox{15cm}{
    \begin{ruledtabular}
    \begin{tabular}{lllccc}
      & \multicolumn{2}{l}{Particles}  & Energy  & Coefficient & State vector  \\ \hline
      Initial state                                                                                          
      & $^3\H_S$                            & $^3\He_D$                                   & $E^{(i)}$       & $c^{(i)}$       & $\ket{\phi^{(i)}}$       \\ \hline
      Intermediate states                                                                                    
      & $^3\He_S^+$, $\bar{\nu}_S$, $e_S^-$ & $^3\He_D$                                   & $E^{(1)}_j$     & $c^{(1)}_j$     & $\ket{\phi^{(1)}_j}$     \\
      & $^3\H_S$                            & $^3\H_D$, $\nu_D$                           & $E^{(2)}_k$     & $c^{(2)}_k$     & $\ket{\phi^{(2)}_k}$     \\
      & $^3\He_S^+$, $\bar{\nu}_S$, $e_S^-$ & $^3\H_D$, $\nu_D$                           & $E^{(3)}_{jk}$  & $c^{(3)}_{jk}$  & $\ket{\phi^{(3)}_{jk}}$  \\
      & $^3\He_S$                           & $^3\H_D$                                    & $E^{(4)}$       & $c^{(4)}$       & $\ket{\phi^{(4)}}$       \\
      & $^3\H_S$                            & $^3\He_D^+$, $\bar{\nu}_D$ $e^-_D$, $\nu_D$ & $E^{(5)}_{kl}$  & $c^{(5)}_{kl}$  & $\ket{\phi^{(5)}_{kl}}$  \\
      & $^3\He_S^+$, $\bar{\nu}_S$, $e_S^-$ & $^3\He_D^+$, $\bar{\nu}_D$ $e^-_D$, $\nu_D$ & $E^{(6)}_{jkl}$ & $c^{(6)}_{jkl}$ & $\ket{\phi^{(6)}_{jkl}}$ \\ \hline
      Final state                                                                                        
      & $^3\He_S$                           & $^3\He_D^+$, $\bar{\nu}_D$ $e^-_D$          & $E^{(f)}_l$     & $c^{(f)}_l$     & $\ket{\phi^{(f)}_l}$
    \end{tabular}
    \end{ruledtabular}}
  \end{center}
  \caption{Classification of the states appearing in a \MB\ neutrino experiment.}
  \label{tab:states}
\end{table}
$^3\H$ and $^3\He$ denote the two types of atoms in the experiment, and the 
index $S$ or $D$ shows whether the respective particle is initially localized 
at the source or at the detector. For those states for which we have written
the electron participating in the reaction and the $^3\He^+$ ions separately,
we imply that the electron may be either free or in an atomic bound state, 
while for the other states only bound electrons are considered. The upper  
index $(i)$ stands for the initial state, the indices $(1)$ through $(6)$ 
denote intermediate states, and $(f)$ stands for the final state, after 
the decay of the source particle, the absorption of the emitted neutrino 
in the detector and the decay of the produced tritium. The lower indices 
stand for the various quantum numbers of the particles; for example,
$j$ encodes the momenta and the spins of $\bar{\nu}_S$ and $e_S^-$, and
the information whether $e_S^-$ is bound or free.

The evolution of the system is governed by the interaction Hamiltonian 
$H_1=\Hamil_S^+ + \Hamil_D^- +\tilde{\Hamil}_D^+ +H.c$, where  
{\allowdisplaybreaks
\begin{align}
  \Hamil_S^+ &= \int \! d^3 x \, \frac{1}{\sqrt{2}} G_F \cos\theta_c 
    \bra{{}^3\He} J^\mu \ket{{}^3\H} \,
                \bar{\psi}_{e,S} \gamma_\mu (1 - \gamma^5) \psi_\nu\,, 
  \label{eq:Hs+} \\
  \Hamil_D^- &= \int \! d^3 x \, \frac{1}{\sqrt{2}} G_F \cos\theta_c 
                    \bra{{}^3\H} J^\mu \ket{{}^3\He} \,
                \bar{\psi}_\nu \gamma_\mu (1 - \gamma^5) \psi_{e,D}\,, 
  \label{eq:Hd-} \\
  \tilde{\Hamil}_D^+ &= \int \! d^3 x \,\frac{1}{\sqrt{2}} G_F \cos\theta_c 
                        \bra{{}^3\He} J^\mu \ket{{}^3\H} \,
                        \bar{\psi}_{e,S} \gamma_\mu (1 - \gamma^5) \psi_\nu\,.
  \label{eq:tildeHd}
\end{align}}
The Hermitian conjugates of these operators will be denoted $\Hamil_S^-$, 
$\Hamil_D^+$ and $\tilde{\Hamil}_D^-$. The Hamiltonians $\Hamil_S^+$ and 
$\tilde{\Hamil}_D^+$ describe tritium decay in the source and detector 
respectively, whereas $\Hamil_D^-$  describes the $\bar{\nu}_e$ capture in the 
detector. Although the Hamiltonians are essentially related by 
$\Hamil_S^+ = \Hamil_D^+ = \tilde{\Hamil}_D^+$, we will treat them as distinct 
operators throughout this appendix to keep our derivation more transparent 
and more general. For the matrix elements of the transitions, the following
relations hold: 
\begin{align}
  \begin{aligned}
    \bra{\phi^{(i)}} \Hamil_S^- \ket{\phi^{(1)}_j}
      &= \bra{\phi^{(2)}_k}    \Hamil_S^- \ket{\phi^{(3)}_{jk}}
       = \bra{\phi^{(5)}_{kl}} \Hamil_S^- \ket{\phi^{(6)}_{jkl}}, \\
    \bra{\phi^{(i)}} \Hamil_D^+ \ket{\phi^{(2)}_k}
      &= \bra{\phi^{(1)}_j}    \Hamil_D^+ \ket{\phi^{(3)}_{jk}}, \\
    \bra{\phi^{(4)}} \tilde{\Hamil}_D^- \ket{\phi^{(f)}_l}
      &= \bra{\phi^{(2)}_k}    \tilde{\Hamil}_D^- \ket{\phi^{(5)}_{kl}}
       = \bra{\phi^{(3)}_{jk}} \tilde{\Hamil}_D^- \ket{\phi^{(6)}_{jkl}},
  \end{aligned}
  \label{eq:A-identities}
\intertext{and similarly,}
  \begin{aligned}
    E^{(i)} - E^{(1)}_j &= E^{(2)}_k - E^{(3)}_{jk}
                         = E^{(5)}_{kl} - E^{(6)}_{jkl}, \\
    E^{(i)} - E^{(2)}_k &= E^{(1)}_j - E^{(3)}_{jk},     \\
    E^{(4)} - E^{(f)}_l &= E^{(2)}_k - E^{(5)}_{kl} 
                         = E^{(3)}_{jk} - E^{(6)}_{jkl}.
  \end{aligned}
  \label{eq:E-identities}
\end{align}
They follow from the fact that the corresponding processes differ
only by the spectator particles. The evolution equations for the system are 
\begin{align}
  i \dot{c}^{(i)}   &=
      \sum_j \bra{\phi^{(i)}} \Hamil_S^- \ket{\phi^{(1)}_j} \, c^{(1)}_j
    + \sum_k \bra{\phi^{(i)}} \Hamil_D^+ \ket{\phi^{(2)}_k} \, c^{(2)}_k,
                                                       \label{eq:ci-dot}\\
  i \dot{c}^{(1)}_j &=
      \bra{\phi^{(1)}_j} \Hamil_S^+ \ket{\phi^{(i)}} \, c^{(i)}
    + \sum_k \bra{\phi^{(1)}_j} \Hamil_D^+ \ket{\phi^{(3)}_{jk}} \, c^{(3)}_{jk},
                                                       \label{eq:c1-dot}\\
  i \dot{c}^{(2)}_k &=
      \bra{\phi^{(2)}_k} \Hamil_D^- \ket{\phi^{(i)}} \, c^{(i)}
    + \sum_j \bra{\phi^{(2)}_k} \Hamil_S^- \ket{\phi^{(3)}_{jk}} \, c^{(3)}_{jk}
    + \sum_l \bra{\phi^{(2)}_k} \tilde{\Hamil}_D^- \ket{\phi^{(5)}_{kl}} \, c^{(5)}_{kl},
                                                       \label{eq:c2-dot}\\
  i \dot{c}^{(3)}_{jk} &=
      \bra{\phi^{(3)}_{jk}} \Hamil_D^- \ket{\phi^{(1)}_j} \, c^{(1)}_j
    + \bra{\phi^{(3)}_{jk}} \Hamil_S^+ \ket{\phi^{(2)}_k} \, c^{(2)}_k
    + \sum_l \bra{\phi^{(3)}_{jk}} \tilde{\Hamil}_D^- \ket{\phi^{(6)}_{jkl}} \, c^{(6)}_{jkl},
                                                       \label{eq:c3-dot}\\
  i \dot{c}^{(4)} &=
      \sum_j \bra{\phi^{(4)}} \Hamil_D^-(\tp) \ket{\phi^{(1)}_j} \, c^{(1)}_j
    + \sum_k \bra{\phi^{(4)}} \Hamil_S^+ \ket{\phi^{(2)}_k} \, c^{(2)}_k
    + \sum_l \bra{\phi^{(4)}} \tilde{\Hamil}_D^- \ket{\phi^{(f)}_l} \, c^{(f)}_{l},
                                                       \label{eq:c4-dot}\\
  i \dot{c}^{(5)}_{kl} &=
      \bra{\phi^{(5)}_{kl}} \tilde{\Hamil}_D^+ \ket{\phi^{(2)}_k} \, c^{(2)}_k
    + \sum_j \bra{\phi^{(5)}_{kl}} \Hamil_S^- \ket{\phi^{(6)}_{jkl}} \, c^{(6)}_{jkl},
                                                       \label{eq:c5-dot}\\
  i \dot{c}^{(6)}_{jkl} &=
      \bra{\phi^{(6)}_{jkl}} \Hamil_S^+ \ket{\phi^{(5)}_{kl}} \, c^{(5)}_{kl}
    + \bra{\phi^{(6)}_{jkl}} \tilde{\Hamil}_D^+ \ket{\phi^{(3)}_{jk}} \, c^{(3)}_{jk},
                                      \phantom{\sum_k} \label{eq:c6-dot}\\
  i \dot{c}^{(f)}_l &=
      \sum_k \bra{\phi^{(f)}} \Hamil_S^+ \ket{\phi^{(5)}_{kl}} \, c^{(5)}_{kl}
    + \bra{\phi^{(f)}_l} \tilde{\Hamil}_D^+ \ket{\phi^{(4)}} \, c^{(4)}.
                                                       \label{eq:cf-dot}
\end{align}
We treat all processes that occur within the source or within the detector 
non-perturbatively, while first-order perturbation theory will be  
used for processes that require the propagation of a neutrino between 
the source and the detector. This second kind of transitions is suppressed
due to the smallness of the solid angle at which the detector is seen from 
the source. Consequently, we include only the respective forward reactions
(i.e.~those proceeding downward in the scheme of Table~\ref{tab:states}),
but neglect the feedback terms, which would otherwise appear in the 
equations for $c^{(1)}_j$, $c^{(2)}_k$, and $c^{(5)}_{kl}$. The feedback 
of $\ket{\phi^{(f)}_l}$ to $\ket{\phi^{(4)}}$ is included because the 
production of both states from the initial state requires a single neutrino 
propagation between the source and the detector. The sums in 
Eqs.~\eqref{eq:ci-dot} -- \eqref{eq:cf-dot} symbolically denote the summation 
over the relevant discrete indices and integration over the continuous 
variables.  

The initial conditions for the equation system \eqref{eq:ci-dot} --
\eqref{eq:cf-dot} are given by  $c^{(i)}(0) = 1$ with all other
coefficients vanishing at $t = 0$.

Our ultimate goal is to solve the evolution equations for $c^{(4)}(t)$,
which determines the $^3\H$ abundance in the detector at time $t$. 
It is convenient to first consider the closed subsystem formed by
Eqs.~\eqref{eq:ci-dot}, \eqref{eq:c1-dot}, \eqref{eq:c2-dot},
\eqref{eq:c3-dot}, \eqref{eq:c5-dot} and \eqref{eq:c6-dot}, which we solve 
from the bottom upwards. We start by integrating Eq.~\eqref{eq:c6-dot}  
to obtain an expression for $c^{(6)}_{jkl}$, which we then insert  
into Eq.~\eqref{eq:c5-dot}. This yields 
\begin{align}
  i \dot{c}^{(5)}_{kl}(t) &=
    \bra{\phi^{(5)}_{kl}} \tilde{\Hamil}_D^+(t) \ket{\phi^{(2)}_k} c^{(2)}_k(t)
                                                       \nonumber\\
  &\hspace{0.5cm}
    - i \sum_j \int_0^t d\tp\,
        \bra{\phi^{(5)}_{kl}} \Hamil_S^-(t) \ket{\phi^{(6)}_{jkl}} \,
        \bra{\phi^{(6)}_{jkl}} \Hamil_S^+(\tp) \ket{\phi^{(5)}_{kl}} \,
        c^{(5)}_{kl}(\tp)
                                                       \nonumber\\
  &\hspace{0.5cm}
    - i \sum_j \int_0^t d\tp\,
        \bra{\phi^{(5)}_{kl}} \Hamil_S^-(t) \ket{\phi^{(6)}_{jkl}} \,
        \bra{\phi^{(6)}_{jkl}} \tilde{\Hamil}_D^+(\tp) \ket{\phi^{(3)}_{jk}} \,
        c^{(3)}_{jk}(\tp)\,.
  \label{eq:c5-dot-1}
\end{align}
Consider first the second term, which describes the effect on 
$\ket{\phi^{(5)}_{kl}}$ of its decay into $\ket{\phi^{(6)}_{jkl}}$. 
Following the Weisskopf-Wigner procedure as described in~\cite{Cohen:QM2},
we split the quantum numbers indexed by $j$ into the energy $E^{(6)}$ 
and the remaining parameters $\beta$. Denoting the density of states 
(the number of states per unit energy interval) by $\rho(E^{(6)}, \beta)$, one 
can make the replacements 
\begin{align}
  \ket{\phi^{(6)}_{jkl}}\rightarrow \ket{\phi^{(6)}_{kl}; E^{(6)}, \beta}\,, 
\qquad\qquad
  \sum_j  \rightarrow \sum_\beta\, \int\!dE^{(6)} 
\rho(E^{(6)}, \beta)
  \label{eq:sum-to-int}
\end{align}
in the second term of Eq.~\eqref{eq:c5-dot-1}, which gives 
\begin{align}
  - i  \int\!dE^{(6)} K(E^{(6)}) \,
        \int_0^t d\tp\, e^{i(E^{(5)}_{kl} - E^{(6)})(t - \tp)} \,
        c^{(5)}_{kl}(\tp)\,.
  \label{eq:c5-dot-2}
\end{align}
Here we have explicitly written down the time dependence of the matrix 
elements and introduced the quantity  
\begin{align}
K(E^{(6)})= \sum_\beta\, \left|\bra{\phi^{(5)}_{kl}} \Hamil_S^-(0) 
\ket{\phi^{(6)}_{kl}; E^{(6)}, \beta}\right|^2\,\rho(E^{(6)}, \beta)\,, 
\label{eq:K}
\end{align}
which is a smooth (non-oscillating) function of energy. More specifically, 
$K(E^{(6)})$ represents a broad bump of width $\mathcal{O}(m_W)$, so that a
non-negligible contribution to the energy integral in~\eqref{eq:c5-dot-2}
can only arise if $t - \tp \lesssim 1/m_W$. Otherwise, the integrand is fast 
oscillating and the integral is strongly suppressed. Therefore, we can to 
a very good accuracy replace $c^{(5)}_{kl}(\tp)$ by  $c^{(5)}_{kl}(t)$ in 
Eq.~\eqref{eq:c5-dot-2} and pull it out of the integral over $\tp$ (we assume 
that $c^{(5)}_{kl}(t)$ is approximately constant over time intervals of 
order $1/m_W$. This assumption will be justified {\it a posteriori} by 
inspecting the obtained expression for $c^{(5)}_{kl}(t)$). For $t \gg 1/m_W$ 
we thus obtain 
\begin{align}
  & - i c^{(5)}_{kl}(t) \int\!dE^{(6)} \, K(E^{(6)})
      \int_0^t d\tp\, e^{i(E^{(5)}_{kl} - E^{(6)})(t - \tp)}
                                                       \nonumber\\
  &\simeq
    -i c^{(5)}_{kl}(t) \int\!dE^{(6)} \bigg[
          \pi \delta(E^{(5)}_{kl} - E^{(6)})
          + i P \bigg(\frac{1}{E^{(5)}_{kl} - E^{(6)}} \bigg)
        \bigg] K(E^{(6)})                              \nonumber\\
&= -i \bigg( \frac{\gamma}{2} + i \delta E \bigg) c^{(5)}_{kl}(t)\,,
  \label{eq:c5-dot-3}
\end{align}
where
\begin{align}
  \gamma   = 2 \pi K(E^{(5)}_{kl})\,, \qquad\qquad
  \delta E = P \int\!dE^{(6)} \frac{K(E^{(6)})}{E^{(5)}_{kl} - E^{(6)}}\,,
  \label{eq:gamma}
\end{align}
and $P$ denotes the principal value. As follows from the definition of the 
function $K(E)$ in Eq.~\eqref{eq:K} and Fermi's golden rule, $\gamma$ 
is just the decay width of $^3\H$ in the source. The quantity $\delta E$ is 
the mass renormalization of the particles forming $\ket{\phi^{(5)}_{kl}}$.
From now on, we will omit $\delta E$ and similar quantities in subsequent 
formulas, assuming that they are already included in the definition of the 
physical masses of the involved particles. The formal solution to 
Eq.~\eqref{eq:c5-dot-1} is 
\begin{align}
  c^{(5)}_{kl}(t) &=
    - i \int_0^t\! d\tp\, \bra{\phi^{(5)}_{kl}} \tilde{\Hamil}_D^+(\tp) 
      \ket{\phi^{(2)}_k} \,
      e^{-\frac{1}{2} \gamma (t - \tp)} \, c^{(2)}_k(\tp) \label{eq:c5-1} \\
  &\hspace{0.5cm}
    + (-i)^2 \sum_j \int_0^t\! d\tp \int_0^{\tp}\! d\tpp
       \bra{\phi^{(5)}_{kl}} \Hamil_S^-(\tp) \ket{\phi^{(6)}_{jkl}} \,
       \bra{\phi^{(6)}_{jkl}} \tilde{\Hamil}_D^+(\tpp) \ket{\phi^{(3)}_{jk}} \,
       e^{-\frac{1}{2} \gamma (t - \tp)} \, c^{(3)}_{jk}(\tpp)\,. \nonumber
\end{align}
By a similar argument, we obtain from Eq.~\eqref{eq:c3-dot}:
\begin{align}
  i \dot{c}^{(3)}_{jk}(t) &=
      \bra{\phi^{(3)}_{jk}} \Hamil_D^-(t) \ket{\phi^{(1)}_j} \, c^{(1)}_j(t)
    + \bra{\phi^{(3)}_{jk}} \Hamil_S^+(t) \ket{\phi^{(2)}_k} \, c^{(2)}_k(t)
    - i \frac{\tilde{\gamma}}{2} c^{(3)}_{jk}(t)
                                                       \nonumber\\
  &\hspace{0.5cm}
    - i \sum_l \int_0^t\! d\tp\,
      \bra{\phi^{(3)}_{jk}} \tilde{\Hamil}_D^-(t) \ket{\phi^{(6)}_{jkl}} \,
      \bra{\phi^{(6)}_{jkl}} \Hamil_S^+(\tp) \ket{\phi^{(5)}_{kl}} \,
      c^{(5)}_{kl}(\tp)\,,
  \label{eq:c3-dot-1}
\end{align}
where the decay width of $^3\H$ in the detector, $\tilde{\gamma}$,
has been defined in analogy with Eq.~\eqref{eq:gamma}. We will now show
that the last term of Eq.~\eqref{eq:c3-dot-1} can be neglected. To this 
end, we insert in it the expression for $c^{(5)}_{kl}(t)$ 
from~\eqref{eq:c5-1}, which yields
\begin{align}
  &\ (-i)^2 \sum_l \int_0^t\! d\tp \int_0^\tp\! d\tpp\,
    |\bra{\phi^{(2)}_k} \tilde{\Hamil}_D^-(0) \ket{\phi^{(5)}_{kl}}|^2 \,
    \bra{\phi^{(6)}_{jkl}} \Hamil_S^+(\tp) \ket{\phi^{(5)}_{kl}} \nonumber\\
  &\hspace{7cm} \cdot
    e^{i \left(E^{(2)}_k - E^{(5)}_{kl}\right)(t - \tpp)} \,
    e^{-\frac{1}{2} \gamma (\tp - \tpp)} \, c^{(2)}_k(\tpp)      \nonumber\\
  +&\ (-i)^3 \sum_l \int_0^t\! d\tp \int_0^\tp\! d\tpp \int_0^\tpp\! d\tppp\,
    |\bra{\phi^{(2)}_k} \tilde{\Hamil}_D^-(0) \ket{\phi^{(5)}_{kl}}|^2 \,
    \bra{\phi^{(6)}_{jkl}} \Hamil_S^+(\tp) \ket{\phi^{(5)}_{kl}}
    \sum_{j^\prime} \bra{\phi^{(5)}_{kl}} \Hamil_S^-(\tpp) \ket{\phi^{(6)}_{j^\prime kl}} \,
                                                                 \nonumber\\
  &\hspace{7cm} \cdot
    e^{i \left(E^{(2)}_k - E^{(5)}_{kl}\right)(t - \tppp)} \, 
    c^{(3)}_{j^\prime k}(\tppp)\,.
  \label{eq:c3-dot-2}
\end{align}
Here we have used Eqs.~\eqref{eq:A-identities} and \eqref{eq:E-identities}. 
We will show now that the first term of Eq.~\eqref{eq:c3-dot-2} can be 
neglected; a similar argument can be used to justify the neglect of the 
second term.  

Replacing the index $l$ by $E^{(5)}$ and $\tilde{\beta}$ in analogy with  
Eq.~\eqref{eq:sum-to-int}, we obtain 
\begin{align}
  (-i)^2 \int\!dE^{(5)}\, \tilde{K}(E^{(5)})
    \int_0^t\! d\tp \int_0^\tp\! d\tpp\,
    \bra{\phi^{(6)}_{jkl}} \Hamil_S^+(\tp) \ket{\phi^{(5)}_{kl}} \,
    e^{i \left(E^{(2)}_k - E^{(5)}\right)(t - \tpp)} \,
    e^{-\frac{1}{2} \gamma (\tp - \tpp)} \, c^{(2)}_k(\tpp)\,,
  \label{eq:c3-dot-3}
\end{align}
with $\tilde{K}(E^{(5)})$ defined analogously to $K(E^{(6)})$. 
As in Eq.~\eqref{eq:c5-dot-2}, the energy integral is non-negligible only
if $t - \tpp \lesssim 1/m_W$. We see immediately that here this condition
also implies $t - \tp \lesssim 1/m_W$. Consequently, we may pull out of the
integral those terms which remain approximately constant over time
intervals $\mathcal{O}(1/m_W)$, which gives 
\begin{align}
  &\ (-i)^2 \bra{\phi^{(6)}_{jkl}} \Hamil_S^+(t) \ket{\phi^{(5)}_{kl}} c^{(2)}_k(t) 
    \int\!dE^{(5)}\, \tilde{K}(E^{(5)}) \int_0^t\! d\tp \int_0^\tp\! d\tpp\,
    e^{i \left(E^{(2)}_k - E^{(5)}\right)(t - \tpp)}   \nonumber\\
  \sim&\ (-i)^2 \bra{\phi^{(6)}_{jkl}} \Hamil_S^+(t) \ket{\phi^{(5)}_{kl}}
    c^{(2)}_k(t) \frac{1}{m_W} \int\!dE^{(5)}\,
    \bigg[ \pi \delta(E^{(2)}_k - E^{(5)})
            + i P \bigg(\frac{1}{E^{(2)}_k - E^{(5)}} \bigg) \bigg]
    \tilde{K}(E^{(5)})                                 \nonumber \\
  \sim&\ \mathcal{O} \bigg( \frac{\tilde{\gamma}}{m_W} \bigg),
  \label{eq:c3-dot-4}
\end{align}
which is negligible compared to the other terms contributing to
$\dot{c}^{(3)}_{jk}(t)$ (cf.~Eq.~\eqref{eq:c3-dot-1}). This result
already suggests the general rule that the only transitions which may
contribute sizeably to the evolution equations are those corresponding to
the direct production of the states (i.e.~production with a minimum number
of intermediate steps), and those corresponding to direct feedback
from a daughter state into its immediate parent state, e.g.~from
$\ket{\phi^{(6)}_{jkl}}$ into $\ket{\phi^{(3)}_{jk}}$. All terms
corresponding to more complicated interaction chains are negligible.
One can now solve Eq.~\eqref{eq:c3-dot-2} for $c^{(3)}_{jk}$: 
\begin{align}
  c^{(3)}_{jk}(t) &=
  -i\int_0^t\! d\tp \bra{\phi^{(3)}_{jk}} \Hamil_D^-(\tp) \ket{\phi^{(1)}_j} \,
        e^{-\frac{1}{2} \tilde{\gamma} (t - \tp)} \, c^{(1)}_j(\tp)
                                                       \nonumber\\
  &\hspace{0.5cm}
  -i \int_0^t\! d\tp \bra{\phi^{(3)}_{jk}} \Hamil_S^+(\tp) \ket{\phi^{(2)}_k}\,
        e^{-\frac{1}{2} \tilde{\gamma} (t - \tp)} \, c^{(2)}_k(\tp)\,.
  \label{eq:c3-1}
\end{align}
Next, we plug our expressions~\eqref{eq:c3-1} and~\eqref{eq:c5-1} for 
$c^{(3)}_{jk}$ and $c^{(5)}_{kl}$ into Eq.~\eqref{eq:c2-dot}:
\begin{align}
  i \dot{c}^{(2)}_k(t) &=
    \bra{\phi^{(2)}_k} \Hamil_D^-(t) \ket{\phi^{(i)}} c^{(i)}(t)
    - i \frac{\gamma}{2} c^{(2)}_k(t)
    - i \frac{\tilde{\gamma}}{2} c^{(2)}_{jk}(t)
                                                       \nonumber\\
  &\hspace{0.5cm}
    - i \sum_j \int_0^t\! d\tp\,
      \bra{\phi^{(2)}_k} \Hamil_S^-(t) \ket{\phi^{(3)}_{jk}} \,
      \bra{\phi^{(3)}_{jk}} \Hamil_D^-(\tp) \ket{\phi^{(1)}_j} \,
      e^{-\frac{1}{2} \tilde{\gamma} (t - \tp)} \, c^{(1)}_j(\tp)\,.
  \label{eq:c2-dot-1}
\end{align}
We have omitted a term containing the product of
$\bra{\phi^{(2)}_k} \tilde{\Hamil}_D^- \ket{\phi^{(5)}_{kl}}$,
$\bra{\phi^{(5)}_{kl}} \Hamil_S^- \ket{\phi^{(6)}_{jkl}}$ and
$\bra{\phi^{(6)}_{jkl}} \tilde{\Hamil}_D^+ \ket{\phi^{(3)}_{jk}}$ 
and thus describing the transition chain
$\ket{\phi^{(3)}_{jk}} \rightarrow \ket{\phi^{(6)}_{jkl}} \rightarrow
\ket{\phi^{(5)}_{kl}} \rightarrow \ket{\phi^{(2)}_k}$, because this
term can be shown to be $\mathcal{O}(\tilde{\gamma} / m_W)$ by an 
argument similar to the one we used for Eq.~\eqref{eq:c3-dot-4}. 
The formal solution to Eq.~\eqref{eq:c2-dot-1} is 
\begin{align}
  c^{(2)}_k(t) &=
    -i \int_0^t\! d\tp\, \bra{\phi^{(2)}_k} \Hamil_D^-(\tp) \ket{\phi^{(i)}} \,
        e^{- \frac{1}{2} \gamma (t - \tp)
           - \frac{1}{2} \tilde{\gamma} (t - \tp)} \, c^{(i)}(\tp)
  \label{eq:c2-1} \\
  &\hspace{-0.5cm}
    + (-i)^2 \sum_j \int_0^t\! d\tp \int_0^\tp\! d\tpp\,
        \bra{\phi^{(2)}_k} \Hamil_S^-(\tp) \ket{\phi^{(3)}_{jk}} \,
        \bra{\phi^{(3)}_{jk}} \Hamil_D^-(\tpp) \ket{\phi^{(1)}_j} \,
        e^{- \frac{1}{2} \gamma (t - \tp)
           - \frac{1}{2} \tilde{\gamma} (t - \tpp)} \, c^{(1)}_j(\tpp)\,.
                              \nonumber
\end{align}
We now proceed to Eq.~\eqref{eq:c1-dot}:
\begin{align}
  i \dot{c}^{(1)}_j(t) &=
    \bra{\phi^{(1)}_j} \Hamil_S^+(t) \ket{\phi^{(i)}} c^{(i)}(t)
                                                       \nonumber\\
  &\hspace{0.5cm}
    - i \sum_k \int_0^t\! d\tp\,
      \bra{\phi^{(1)}_j} \Hamil_D^+(t) \ket{\phi^{(3)}_{jk}} \,
      \bra{\phi^{(3)}_{jk}} \Hamil_D^-(\tp) \ket{\phi^{(1)}_j} \,
      e^{-\frac{1}{2} \tilde{\gamma} (t - \tp)} \, c^{(1)}_j(\tp)\,.
  \label{eq:c1-dot-1}
\end{align}
The contributions coming from $c^{(2)}_k$ through the transition chain
$\ket{\phi^{(2)}_k} \rightarrow \ket{\phi^{(3)}_{jk}} \rightarrow
\ket{\phi^{(1)}_j}$ are again omitted as being 
$\mathcal{O}(\tilde{\gamma}/m_W)$. The term containing
$\bra{\phi^{(1)}_j} \Hamil_D^+(t) \ket{\phi^{(3)}_{jk}} \,
\bra{\phi^{(3)}_{jk}} \Hamil_D^-(\tp) \ket{\phi^{(1)}_j}$ describes
the direct feedback from $\ket{\phi^{(3)}_{jk}}$ to $\ket{\phi^{(1)}_j}$, 
but since the transition $\ket{\phi^{(1)}_j} \to \ket{\phi^{(3)}_{jk}}$ 
does not occur spontaneously, the corresponding decay width is zero. Indeed, 
when applying the Weisskopf-Wigner procedure, 
we see that the resulting $\delta$-function under the energy integral 
is zero for all allowed energies. 
Thus, the second term in Eq.~\eqref{eq:c1-dot-1} is negligible, and 
the equation is solved by 
\begin{align}
  c^{(1)}_j(t) &=
    - i \int_0^t\! d\tp\,
      \bra{\phi^{(1)}_j} \Hamil_S^+(\tp) \ket{\phi^{(i)}} \,
      c^{(i)}(\tp)\,.
  \label{eq:c1-1}
\end{align}
We can insert this expression, together with $c^{(2)}_k(t)$ from
Eq.~\eqref{eq:c2-1}, into the equation for $c^{(i)}(t)$, and find
\begin{align}
  c^{(i)}(t) = e^{-\frac{1}{2} \gamma t}\,,
  \label{eq:ci-1}
\end{align}
up to a term suppressed by $\tilde{\gamma} / m_W$.
The closed-form expressions for $c^{(1)}_j(t)$, $c^{(2)}_k(t)$,
$c^{(3)}_{jk}(t)$, $c^{(5)}_{kl}(t)$, and $c^{(6)}_{jkl}(t)$
are then
{\allowdisplaybreaks
\begin{align}
  c^{(1)}_j(t) &=
    - i \int_0^t\! d\tp\,
      \bra{\phi^{(1)}_j} \Hamil_S^+(\tp) \ket{\phi^{(i)}} \,
      e^{-\frac{1}{2} \gamma \tp}\,,
  \label{eq:c1-2}                                      \\
  c^{(2)}_k(t) &=
    - i \int_0^t\! d\tp\, \bra{\phi^{(2)}_k} \Hamil_D^-(\tp) \ket{\phi^{(i)}} \,
      e^{- \frac{1}{2} \gamma t
         - \frac{1}{2} \tilde{\gamma} (t - \tp)}\,,
  \label{eq:c2-2}                                      \\
  c^{(3)}_{jk}(t) &=
    (-i)^2 \bigg[ \int_0^t\! d\tp\,
        \bra{\phi^{(1)}_j} \Hamil_S^+(\tp) \ket{\phi^{(i)}} \,
        e^{-\frac{1}{2} \gamma \tp} \bigg]
      \bigg[ \int_0^t\! d\tp\,
        \bra{\phi^{(2)}_k} \Hamil_D^-(\tp) \ket{\phi^{(i)}} \,
        e^{-\frac{1}{2} \tilde{\gamma} (t - \tp)} \bigg]\,,
  \label{eq:c3-2}                                      \\
  c^{(5)}_{kl}(t) &=
    (-i)^2 \int_0^t\! d\tp \int_0^\tp\! d\tpp\,
      \bra{\phi^{(5)}_{kl}} \tilde{\Hamil}_D^+(\tp) \ket{\phi^{(2)}_k} \,
      \bra{\phi^{(2)}_k} \Hamil_D^-(\tpp) \ket{\phi^{(i)}} \,
      e^{- \frac{1}{2} \gamma t
         - \frac{1}{2} \tilde{\gamma} (\tp - \tpp)}\,,
  \label{eq:c5-2}                                      \\
  c^{(6)}_{jkl}(t) &=
    (-i)^3 \bigg[ \int_0^t\! d\tp\,
      \bra{\phi^{(1)}_j} \Hamil_S^+(\tp) \ket{\phi^{(i)}} \,
      e^{-\frac{1}{2} \gamma \tp} \bigg]
                                                       \nonumber\\
  &\hspace{2cm} \cdot
    \bigg[ \int_0^t\! d\tp\! \int_0^\tp\! d\tpp\,
      \bra{\phi^{(5)}_{kl}} \tilde{\Hamil}_D^+(\tp) \ket{\phi^{(2)}_k} \,
      \bra{\phi^{(2)}_k} \Hamil_D^-(\tpp) \ket{\phi^{(i)}} \,
      e^{- \frac{1}{2} \tilde{\gamma} (\tp - \tpp)} \bigg]\,.
  \label{eq:c6-1}
\end{align}}
In the expressions for $c^{(3)}_{jk}(t)$ and $c^{(6)}_{jkl}(t)$, we have
used the identity
\begin{align}
  \int_0^t\! d\tp \int_0^\tp\! d\tpp = \int_0^t\! d\tpp \int_\tpp^t\! d\tp.
\end{align}
Eqs.~\eqref{eq:ci-1} -- \eqref{eq:c6-1} show that all coefficients are
slowly varying over time intervals of order $1/m_W$, which provides the
{\it a posteriori} justification for pulling them out of the time integrals
when applying the Weisskopf-Wigner procedure.

We have now all the ingredients required to solve for $c^{(4)}(t)$.
We insert Eqs.~\eqref{eq:c1-2}, \eqref{eq:c2-2} and~\eqref{eq:cf-dot} into 
Eq.~\eqref{eq:c4-dot}, neglect the $\mathcal{O}(\tilde{\gamma}/m_W)$ 
contribution from the reaction chain $\ket{\phi^{(5)}_{kl}} \rightarrow
\ket{\phi^{(f)}_l} \rightarrow \ket{\phi^{(4)}}$, and apply the 
completeness relations 
\begin{align}
  \sum_j \ket{\phi^{(1)}_j} \bra{\phi^{(1)}_j} = 1\,,\qquad\quad
  \sum_k \ket{\phi^{(2)}_k} \bra{\phi^{(2)}_k} = 1
\end{align}
to dispose of the sums over $j$ and $k$ and of the intermediate
bra- and ket-vectors in the products of matrix elements. This leads
us to the main result of this appendix,
\begin{align}
  c^{(4)}(t) &= (-i)^2 \int_0^t\! d\tp \int_0^\tp\! d\tpp \,
    \bra{\phi^{(4)}} \Big[
        \Hamil_D^-(\tp)  e^{-\frac{1}{2} \tilde{\gamma} (t - \tp)} \,
        \Hamil_S^+(\tpp) e^{-\frac{1}{2} \gamma \tpp}    \nonumber\\
  &\hspace{5cm}
      + \Hamil_S^+(\tp)  e^{-\frac{1}{2} \gamma \tp} \,
        \Hamil_D^-(\tpp) e^{-\frac{1}{2} \tilde{\gamma} (t - \tpp)}
    \Big] \ket{\phi^{(i)}}\,.
  \label{eq:c4-3}
\end{align}
We see that $c^{(4)}(t)$ is given by the time-ordered product of the
two interaction Hamiltonians, supplemented by the classically expected
exponential decay factors. After inserting the appropriate expressions
for $\Hamil_S^+$ and $\Hamil_D^-$, finally setting $\tilde{\gamma} = \gamma$ 
and applying the Feynman rules, Eq.~\eqref{eq:c4-3} leads directly to
Eq.~\eqref{eq:QFT-A5} of Sec.~\ref{sec:QFT-nat}.

For completeness, we also give the expression for $c^{(f)}_l(t)$:
\begin{align}
  c^{(f)}_l(t) &=
    (-i)^3 \int_0^t\! d\tp \int_0^\tp\! d\tpp \int_0^\tpp\! d\tppp
    \bra{\phi^{(f)}_l} \Big[
        \tilde{\Hamil}_D^+(\tp) \,
        \Hamil_D^-(\tpp)  e^{-\frac{1}{2} \tilde{\gamma} (\tp - \tpp)} \,
        \Hamil_S^+(\tppp) e^{-\frac{1}{2} \gamma \tppp}    \nonumber\\
  &\hspace{5cm}
      + \tilde{\Hamil}_D^+(\tp) \,
        \Hamil_S^+(\tpp)  e^{-\frac{1}{2} \gamma \tpp} \,
        \Hamil_D^-(\tppp) e^{-\frac{1}{2} \tilde{\gamma} (\tp - \tppp)} \,
                                                           \nonumber\\
  &\hspace{5cm}
      + \Hamil_S^+(\tp)  e^{-\frac{1}{2} \gamma \tp} \,
        \tilde{\Hamil}_D^+(\tpp)
        \Hamil_D^-(\tppp) e^{-\frac{1}{2} \tilde{\gamma} (\tpp - \tppp)}
    \Big] \ket{\phi^{(i)}}\,.
  \label{eq:cf-1}
\end{align}

Note that an alternative way of solving Eqs.~\eqref{eq:ci-dot}
-- \eqref{eq:cf-dot} is to exploit the fact that, in the closed system
formed by Eqs.~\eqref{eq:ci-dot}, \eqref{eq:c1-dot}, \eqref{eq:c2-dot},
\eqref{eq:c3-dot}, \eqref{eq:c5-dot}, and \eqref{eq:c6-dot}, the processes
in the source and those in the detector can be separated by using a
product ansatz for the coefficients $c$. Once this subsystem is solved,
$c^{(4)}$ and $c^{(f)}_l$ can be computed as above.



\end{document}